\documentclass[aps,prc,reprint,superscriptaddress,floatfix,nofootinbib]{revtex4-2}

\usepackage{amsmath}
\usepackage{graphicx}
\usepackage{dcolumn}
\usepackage{bm}
\usepackage{multirow}
\usepackage{booktabs}
\usepackage{physics}
\usepackage{xcolor}
\usepackage{makecell}
\usepackage[version=4]{mhchem}
\usepackage{hyperref}

\begin{document}


\title{Uncertainty Quantification of Mass Models Using Ensemble Bayesian Model Averaging}



\author{Yukiya Saito}
 \email{yukiya@alum.ubc.ca}
\affiliation{TRIUMF, 4004 Wesbrook Mall, Vancouver, BC V6T 2A3, Canada}%
\affiliation{Department of Physics and Astronomy, The University of British Columbia, Vancouver, BC V6T 1Z1, Canada}%
\affiliation{Department of Physics, University of Notre Dame, Notre Dame, IN 46556, USA}
\affiliation{\textcolor{black}{Department of Physics and Astronomy, University of Tennessee, Knoxville 37996, Tennessee, USA }}

\author{I. Dillmann}
\affiliation{TRIUMF, 4004 Wesbrook Mall, Vancouver, BC V6T 2A3, Canada}
\affiliation{Department of Physics and Astronomy, University of Victoria, Victoria, BC V8P 5C2, Canada}

\author{R. Kr{\"u}cken}
\affiliation{TRIUMF, 4004 Wesbrook Mall, Vancouver, BC V6T 2A3, Canada}
\affiliation{Department of Physics and Astronomy, The University of British Columbia, Vancouver, BC V6T 1Z1, Canada}
\affiliation{Lawrence Berkeley National Laboratory,
Berkeley, California 94720, USA}

\author{M. R. Mumpower}
\affiliation{Theoretical Division, Los Alamos National Laboratory, Los Alamos, NM 87545, USA}
\affiliation{Center for Theoretical Astrophysics, Los Alamos National Laboratory, Los Alamos, NM 87545, USA}

\author{R. Surman}
\affiliation{Department of Physics, University of Notre Dame, Notre Dame, IN 46556, USA}


\date{\today}

\begin{abstract}
Developments in the description of the masses of atomic nuclei have led to various nuclear mass models that provide predictions for masses across the whole chart of nuclides. These mass models play an important role in understanding the synthesis of heavy elements in the rapid neutron capture ($r$-) process. However, it is still a challenging task to estimate the size of uncertainty associated with the predictions of each mass model. In this work, a method called \textit{ensemble Bayesian model averaging} (EBMA) \textcolor{black}{is  introduced to quantify the uncertainty of one-neutron separation energies ($S_{1n}$) which are directly relevant in the calculations of $r$-process observables.} This Bayesian method provides a natural way to perform model averaging, selection, and uncertainty quantification, by combining the mass models as a mixture of normal distributions whose parameters are optimized against the experimental data, employing the Markov chain Monte Carlo (MCMC) method using the No-U-Turn sampler (NUTS). \textcolor{black}{The EBMA model optimized with all the experimental $S_{1n}$ from the AME2003 nuclides are shown to provide reliable uncertainty estimates when tested with the new data in the AME2020. Furthermore, the model has been shown to detect the anomalous behavior in nuclear masses known as the Wigner effect, suggesting the possibility of detecting further deficiencies of theoretical models in light of new experimental data.}
\end{abstract}


\maketitle

\section{\label{sec:Intro}Introduction}
Since the first introduction of the nuclear liquid drop model, the theoretical description of nuclear masses has seen great progress, which gave rise to many related but different approaches. It is now possible to describe the ground state properties of nuclei across the chart of nuclei with theories of different scales: Macroscopic-microscopic theories such as the Finite-Range Droplet Model (FRDM) \cite{Moller1995,Moller2016}, Weizs\"{a}cker-Skyrme (WS) models \cite{Wang2010a,Wang2010b,Liu2011,Wang2014}, microscopically inspired Duflo-Zucker models \cite{Duflo1995}, and more microscopic theories such as nuclear density functional theory (DFT) with different interactions or energy density functionals (EDFs) \cite{Erler2012,Goriely2009,Goriely2016}. 

\textcolor{black}{
Among the theoretical models that describe various aspects of nuclear structure, reactions, and decays,
} 
global mass models play an important role in understanding the origin of heavy elements in the Universe via the rapid neutron capture ($r$-) process \cite{MendozaTemis2015, Martin2016, Mumpower2016}. 
This is because the nuclear masses determine the $Q$-value (energy release \textcolor{black}{or absorption}) of nuclear reactions and decays and \textcolor{black}{
are always required for the calculation of the relevant rates.
} 
The masses of the vast majority of neutron-rich nuclei relevant to the $r$-process have yet to be experimentally studied, \textcolor{black}{therefore, theoretical predictions of the masses must be used instead.}
This means that the mass models used in nucleosynthesis studies may have a significant impact on \textcolor{black}{the prediction of} abundance patterns and kilonova lightcurves~\cite{Zhu2021, Barnes2021}.

\begin{figure*}
\centering
\includegraphics[width=0.8\linewidth]{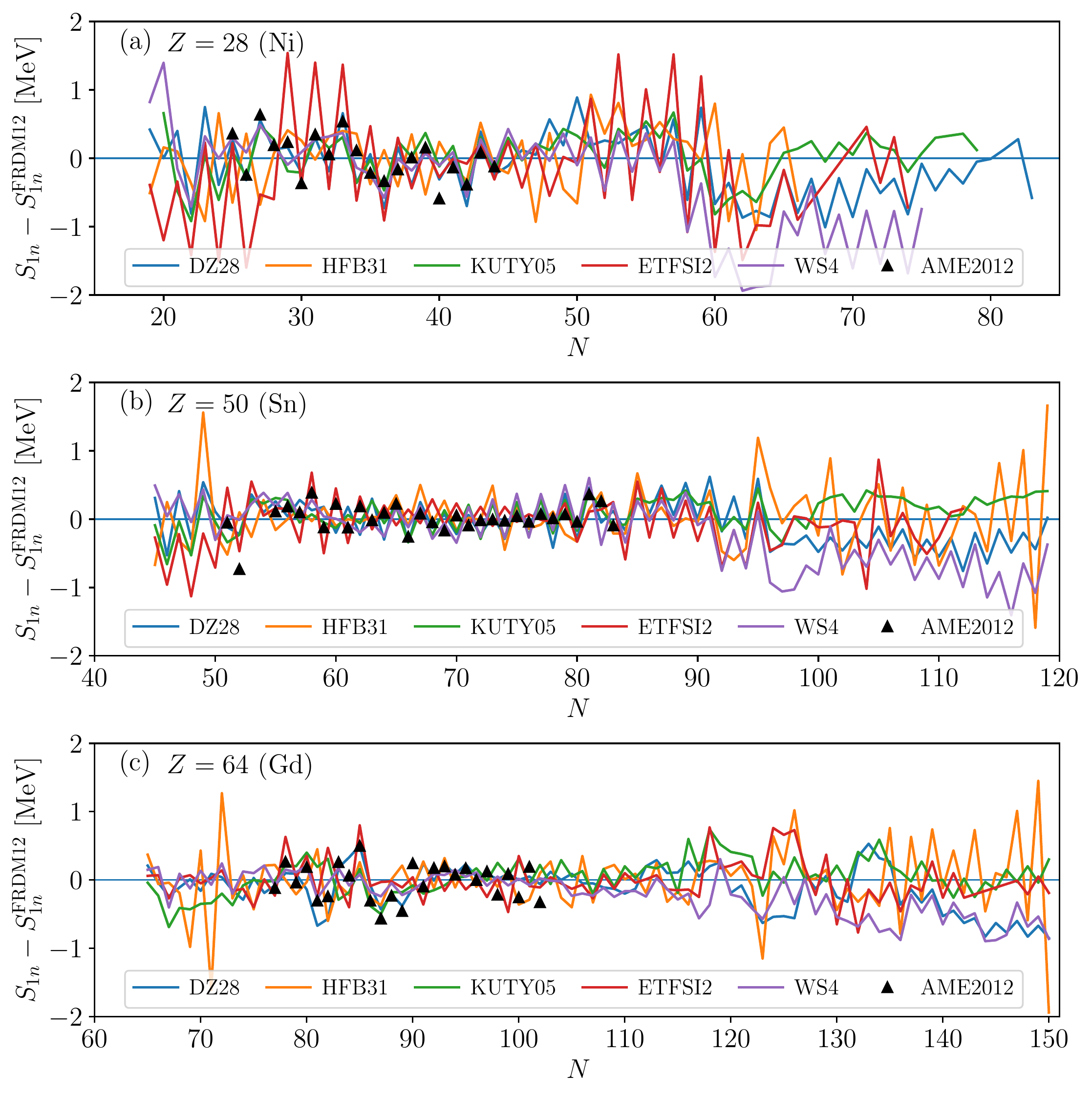}
\caption{\label{fig:Sn-models-comparison}Comparison of one-neutron separation energies ($S_{1n}$) predicted by each mass model used in this study and the experimental masses from the AME2020~\cite{Wang2021}, relative to the predictions of the FRDM2012 for (a) $Z=28$ (Ni), (b) $Z=50$ (Sn), and (c) $Z=64$ (Gd) isotopes. The mass models and references are listed in Section~\ref{sec:SetupNumericalExp}}.
\end{figure*}

One of the challenges in understanding the impact of mass models on nucleosynthesis is that, in general, uncertainty estimates associated with the theoretical masses are not available. Although there has been an effort to quantify the uncertainty in microscopic theories \cite{Dobaczewski2014, McDonnell2015}, the mass models that are commonly used in nucleosynthesis studies, especially macroscopic-microscopic and phenomenological models, do not come with a quantified prediction uncertainty. The root mean square (RMS) error of each mass model can be calculated with respect to the observations, but it most likely underestimates the uncertainty where there is no data (see Figure~\ref{fig:Sn-models-comparison}). 
This poses a challenge in quantifying the uncertainty in the $r$-process nucleosynthesis that arises from uncertain nuclear masses.

As the next-generation radioactive isotope beam facilities allow us to gain access to more neutron-rich isotopes, it will become possible to test the performances of mass models in extremely neutron-rich regions of the chart of nuclides. 
\textcolor{black}{
It would be ideal to have a method that can continuously incorporate new experimental results, test the agreement to theoretical predictions, and update their associated uncertainties in extrapolation to further neutron-rich regions.
}


In this work, we will apply a method called \emph{ensemble Bayesian model averaging} (EBMA) introduced by Ref.~\cite{Raftery2005} to combine available experimental data and multiple theoretical mass models, as well as to quantify the mass uncertainty.
This method models an ensemble of theoretical mass models as a mixture of normal distributions, whose parameters are estimated based on the observations. 
EBMA combines model selection, averaging, and uncertainty quantification in a single framework. 
Although we focus on \textcolor{black}{theoretical one-neutron separation energies ($S_{1n}$)} in this work, this Bayesian method is quite general and can potentially be applied to other nuclear physics observables. 
\textcolor{black}{
We note, however, that the application of this method to quantities such as reaction and decay rates relevant to the $r$-process will require further investigations, due to the smaller number of models available and the fact that the rates can vary by orders of magnitudes between nuclei.
}

Recently, data-driven modeling of nuclear masses using machine learning techniques has gained popularity \cite{Niu2022, Niu2018, Neufcourt2018, Neufcourt2019, Lasseri2020, Utama2016, Lovell2022, Mumpower2022, Mumpower2023}. Probabilistic models have especially achieved high accuracy while providing uncertainty estimates. 
Although there have been attempts to construct physically interpretable models \cite{Mumpower2022}, 
\textcolor{black}{it is generally challenging to gain insight into the underlying physics from machine learning models.}
Nevertheless, the advantages of machine learning models are that they can be created rapidly and often achieve similar performance to state-of-the-art theoretical models. 
\textcolor{black}{
Although they may not necessarily be able to predict new or unknown physics, the flexibility of the models may allow us to combine known physics in potentially novel ways that are difficult to produce through standard modeling.
}

The purpose of this study is not to create another mass model or to improve existing ones with machine learning. Rather, the aim is to investigate how well an ensemble of theoretical models can reproduce experimental data and quantify the performance of each model in the ensemble. This also quantifies the uncertainty in extrapolating the experimental data. Our approach should be considered as a method for model averaging, selection, and uncertainty quantification, using only existing theoretical models.

\textcolor{black}{
This paper is divided into the following sections: in Sec.~\ref{sec:Method}, we discuss the details of the EBMA method and the numerical experiment where we construct EBMA models; in Sec.~\ref{sec:ResultsDiscussion}, we discuss the results of the different approaches for constructing EBMA models and the details of the quantified uncertainties for one neutron separation energies ($S_{1n}$); finally, we summarize the work presented in the paper and describe possible future developments in Sec.~\ref{sec:conclusion}. 
}

\section{\label{sec:Method}Method}
\subsection{\label{subsec:BMA}Bayesian model averaging}
Bayesian model averaging (BMA) is applicable when more than one statistical model that describes the data reasonably well is available, and one wishes to account for the uncertainty in the analysis arising from conditioning on a single model. BMA computes a weighted average of the probability density functions (PDFs), weighted by the posterior probability of the ``correctness'' of each model given the training data. Following the description in Refs.~\cite{Hoeting1999, Raftery2005}, the posterior distribution of the observable of interest $\Delta$ \textcolor{black}{(in our case, it correponds to one-neutron separation energies $S_{1n}$)}, defined by BMA, is
\begin{equation}
    p(\Delta \mid D) = \sum_{k=1}^{K}\; p(\Delta \mid M_k, D)\; p(M_k \mid D),
\end{equation}
where $p(\Delta \mid M_k, D)$ is the posterior PDF of the observable of interest based on a single statistical model $M_k$, and $p(M_k \mid D)$ is the corresponding posterior model probability, which 
represents how well the model $M_k$ fits the data $D$. The posterior model probabilities can be considered as weights, since their sum is equal to 1.

\subsection{\label{subsec:EBMA}Ensemble Bayesian model averaging}
One of the limitations in the applicability of the BMA method is that the participating models themselves must be probabilistic. In nuclear physics, most models are not probabilistic. Therefore, we need to extend the BMA framework to handle such models. \citeauthor{Raftery2005}~\cite{Raftery2005} introduced the \emph{ensemble Bayesian model averaging} (EBMA) method, which computes the weighted average of an ensemble of bias-corrected models, as a finite mixture of normal distributions. In the EBMA framework, the predictive model is 
\begin{equation}
    p(\Delta \mid m_1, \ldots, m_K) = \sum_{k=1}^{K}\; w_k \; g_k(\Delta \mid m_k),
\end{equation}
where \textcolor{black}{$\Delta$ is again the quantity of interest (in our case $S_{1n}$ of some nucleus),} $w_k$ is the weight of \textcolor{black}{the model $k$}, whose posterior represents the probability of the model $k$ being the best one, based on the observed data $D$. \textcolor{black}{Model prediction $m_k$ can be a vector or scalar. In our case, it is a vector of theoretical $S_{1n}$ values predicted by the mass model $k$}.
\textcolor{black}{Since the size of the weight represents the posterior probability of the model being the best one, even if the ensemble includes an extremely inaccurate model, its effect on the predictive distributions of EBMA will be quite small, since an extremely small weight would be assigned to the model.}
$g_k(\Delta \mid m_k)$ is a normal PDF with its mean defined by the \textcolor{black}{model prediction $m_k$} and the standard deviation $\sigma_k$:
\begin{equation}
    g_k(\Delta \mid m_k) = N(\Delta \mid m_k, \sigma_k^2). \label{eq:NormalPredictive} 
\end{equation}
\textcolor{black}{
Although one needs to be cautious of overfitting, the mean of the normal PDF $m_k$ may be replaced by the bias-corrected model predictions, which are discussed in more detail in the following section.
}
In the original EBMA by \citeauthor{Raftery2005}~\cite{Raftery2005}, a constant standard deviation was used across all the models in the ensemble; however, we take it as model dependent (denoted by the subscript $k$), which is a more natural way to construct a mixture model.

\subsubsection{Bias correction}
\textcolor{black}{
In constructing EBMA models, although not strictly necessary, Ref.~\cite{Raftery2005} suggests linearly correcting the bias in the prediction of each model, prior to Bayesian inference of weights and standard deviations. This means replacing $m_k$ in Eq.~\ref{eq:NormalPredictive} with  $a_k + b_k m_k$, where $a_k$ is the intercept coefficient, $b_k$ is the slope coefficient, and $m_k$ is the prediction of the model $k$. In our case, this corresponds to linearly correcting a vector of theoretical $S_{1n}$ values predicted by each mass model.
The original prediction of the model $k$ corresponds to $a_k = 0$ and $b_k = 1$.
}

\textcolor{black}{
Although the bias correction may provide one way to fine-tune the predictions of theoretical models (for more sophisticated approaches of correcting mass models, see e.g. Ref.\cite{Neufcourt2018}) in case the model was constructed based on an older set of experimental data, or the predictions for the nuclei of interest are known to systematically deviate from the experimental results, we note that the use of bias correction can lead to overfitting. We investigate the effect of bias correction in Section~\ref{subsec:biascorrected}. Throughout the manuscript, we do not apply bias correction unless explicitly noted otherwise.
}

\subsubsection{Bayesian inference}
The parameters of interest in our statistical inference are the weights $w_k\;(k=1,\ldots,K)$ and the standard deviations $\sigma_k\;(k=1,\ldots,K)$ of the normal distributions that correspond to each of the theoretical mass models in the ensemble. Therefore, prior distributions for these parameters must be specified. In general, we try to choose the prior distributions to be as weakly informative as possible. For the weights, since they have to sum up to one $\left (\sum_{k=1}^K w_k = 1 \right)$, we model the parameters with a Dirichlet distribution of order $K$, which naturally meets this requirement. 
\textcolor{black}{
Dirichlet distribution of order $K$ has hyperparameters $\alpha_k$ ($k=1,\ldots,K$), thus, the prior distribution is expressed as
\begin{align}
    p&(w_1, w_2,\; \ldots,\; w_K)\nonumber \\
    &= \mathrm{Dirichlet}(w_1, w_2,\; \ldots,\; w_K \mid \alpha_1, \alpha_2,\; \ldots,\; \alpha_K) \label{eq:Dirichlet}.
\end{align}}
\textcolor{black}{
We set all the concentration parameters of the Dirichlet distribution to 1, which makes the prior distribution uniform for all weights $w_1, \ldots, w_K$, to represent the belief that we do not know which model would perform best.}
The prior distributions for the standard deviations are chosen to be exponential distributions with the rate parameters equal to 1, which has been suggested to be one of the weaker priors \cite{Gelman_prior_2017}.

The likelihood of the normal mixture model is defined as
\begin{align}
        L&(w_1,\; \ldots,\; w_K, \; \sigma_1^2,\; \ldots,\; \sigma_K^2) \nonumber\\
        &= \prod_{(N,Z)}\left( \sum_{k=1}^K w_k g_k(\Delta_{(N,Z)} \mid m_{k,(N,Z)})\right),\label{eq:EBMALikelihood}
\end{align}
where \textcolor{black}{$\Delta$ is again the quantity of interest (in our case $S_{1n}$), and $m_k$ is the model predictions (a vector of $S_{1n}$ predicted by the mass model $k$)}. The subscript $(N,Z)$ represents pairs of neutron number $N$ and proton number $Z$ of the nuclei where experimental values exist. In practice, the logarithm of likelihood (log-likelihood) is often used for computation to avoid numerical problems.

With the prior distributions and the likelihood function, it is now possible to formulate the posterior distributions for the parameters of the EBMA model.
\begin{align}
    p(\bm{w}, \bm{\sigma^2} \mid D) \propto L(\bm{w}, \bm{\sigma^2}) \;p(\bm{w})\;p(\bm{\sigma^2}),
\end{align}
where $\bm{w}=w_1, \ldots, w_K$, $\bm{\sigma^2}=\sigma_1^2, \ldots, \sigma_K^2$, and $D$ denotes observational (experimental) data. 
The prior distributions are denoted as $p(\bm{w})$ and $p(\bm{\sigma}^2)$, respectively.

\begin{figure}
\centering
\includegraphics[width=\linewidth]{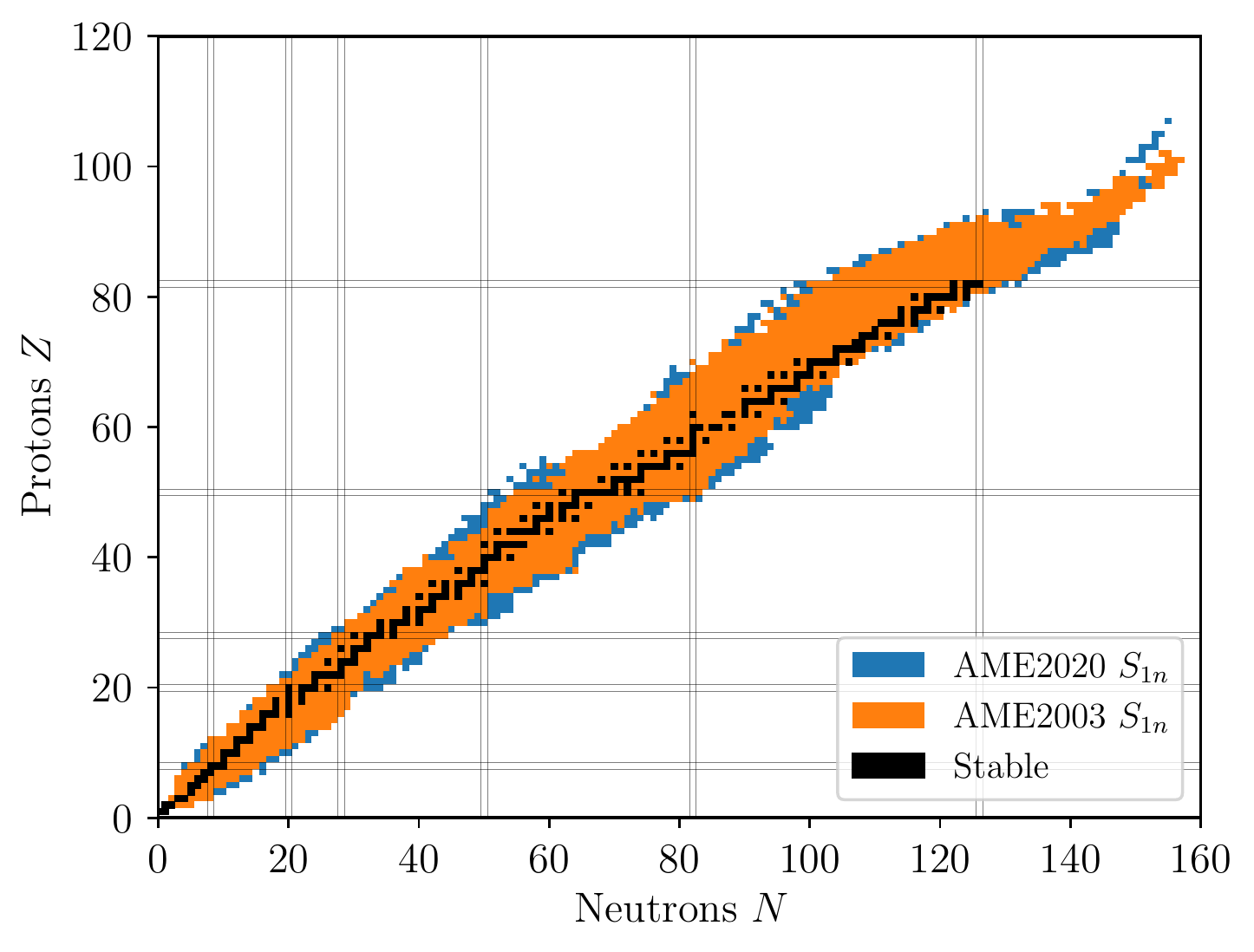}
    \caption{\label{fig:Sn-AME2003_AME2020}Comparison of the AME2003 data with the latest AME2020 data for one-neutron separation energies $S_{1n}$, illustrated on the chart of nuclides. The blue squares show the new $S_{1n}$ data in the AME2020 that did not exist in the AME2003. The $S_{1n}$ values listed in the AME2003 are shown in orange color, aside from the stable nuclides, shown in black. }
\end{figure}

\subsubsection{Predictive variance}
In EBMA models, the uncertainty of the quantity of interest can be interpreted in the form of variance of the posterior predictive distribution. Based on Ref.~\cite{Raftery2005} but reflecting the fact that our $\sigma_k$ depends on model $k$, the predictive variance can be written as 
\begin{align}
    \mathrm{Var}&(\Delta \mid m_1, m_2, \ldots, m_K) \nonumber\\
    =& \sum_{k=1}^K w_k \left( (m_k) -\sum_{i=1}^K w_i (m_i) \right)^2 \nonumber \\
    &+ \sum_{k=1}^K w_k \sigma_k^2, \label{eq:predvar}
\end{align}
where the first term corresponds to the spread of predictions by the member mass models of the ensemble, and the second term corresponds to the expected deviation from the observations of each mass model, weighted by the posterior weights. \textcolor{black}{If bias-corrected models predictions are used, $m_k$ is replaced by $a_k + b_k m_k$. }

\subsubsection{Differences to related works and discussion of models}
It is worth discussing the key differences between our framework and related studies that use the BMA method, namely Refs.\cite{Neufcourt2018, Neufcourt2019, Hamaker2021, Neufcourt2020, Neufcourt2020b}. In their BMA framework, the uncertainty quantification of the considered mass models is performed by constructing Gaussian Process (GP) emulators, which learn the corrections to the mass models from the residuals with respect to the observed values. Therefore, the quality of the prediction and the corresponding uncertainty mainly depend on the performance of the GP emulator. The BMA weights are calculated either based on some criteria such as nuclei being bound or the performances of each mass model on the test data. One of the drawbacks of this method is that the derived weights are point estimates, and the resulting BMA uncertainty is a deterministic weighted average of the GP uncertainties. Furthermore, one has to be cautious when performing extrapolations using GPs, since an unconstrained GP converges to its mean with fixed uncertainty away from the data \cite{Rasmussen2005, Yoshida2020}.

On the other hand, the EBMA framework keeps the point predictions of the mass models in the ensemble. Instead, the weights and variances associated with each mass model are modeled probabilistically based on the experimental data. The probabilistic distributions are reflected onto the resulting predictive uncertainty through the Bayesian framework. \textcolor{black}{This framework directly uses} the predictions of each mass model that constitute the EBMA model; therefore, the local trend of the predictions remains unchanged.

One of the shortcomings of the current method is that inference of posterior weights is performed assuming that all observed data points are equally relevant. In the case of uncertainty quantification of mass models for neutron-rich nuclei, for example, one may wish to estimate the weights by focusing on the data for neutron-rich nuclei. However, this may pose a trade-off since the weights are better estimated using all available data, while only using data in a specific region may better capture the local performances of the mass models. The concept of such location- (input domain-) dependent weights is referred to as ``Bayesian Model Mixing'', put forward by Refs.~\cite{Phillips2021BAND, Semposki2022BMM}.

Further technical development would be required to incorporate location-dependent weights into the averaging of mass models, which will be investigated in the future. In this work, we provide a general methodology for averaging nuclear mass models. The dependence of uncertainty on location is assumed to be represented by the spread of the predictions of different mass models, as shown in Fig.~\ref{fig:Sn-models-comparison}.

\subsection{\label{sec:SetupNumericalExp}Setup of a numerical experiment}
In the numerical experiments discussed in the current work, all probabilistic models have been implemented using PyMC \cite{Salvatier2016}, which is a probabilistic programming language written in Python. PyMC offers an implementation of a highly efficient sampler called No-U-Turn-Sampler (NUTS), which adaptively tunes the parameters associated with the Hamiltonian (or Hybrid) Monte Carlo method \cite{Neal_BNN_2012, Neal2011}. Conventionally, parameter estimation in mixture models is performed with the Expectation Maximization (EM) algorithm to avoid the so-called ``label switching problem'' \cite{Stephens2000, Jasra2005}. The label switching problem arises in mixture models such as EBMA models, since the likelihood (Eq.~\ref{eq:EBMALikelihood}) remains unchanged under permutation of the labels ($k=1,\; \ldots\; K$) of the mixture components $g_k(\Delta \mid m_k)$. This makes the analysis of the posterior distributions challenging. Although, the EM algorithm does not guarantee convergence to the global optimal weights and variances, especially in high-dimensional problems. 

Furthermore, MCMC methods would be able to provide much more complete information on the posterior distributions. In our numerical experiments, we did not find evidence of a label switching problem due to employing the MCMC method.
This is most likely because, in our normal mixture models, the means of the normal distributions are always specified by the predictions of bias-corrected mass models, which works as an identifiability constraint.

The quantity of interest in our study is the one-neutron separation energy ($S_{1n}$), which is directly relevant to the $r$-process. \textcolor{black}{This is because} in nucleosynthesis calculations (post-processing of hydrodynamical simulations), photodissociation rates (denoted as $\lambda_{(\gamma, n)}$ below), \textcolor{black}{for example,} are calculated from the neutron capture rate via detailed balance:
\begin{align}
    \lambda_{(\gamma,n)} = &\expval{\sigma v}_{(n,\gamma)} \cdot \frac{G(N,Z)\cdot G(1,0)}{G(N+1,Z)} \cdot \left( \frac{A}{A+1} \right)^{3/2} \nonumber \\
    &\cdot \left( \frac{m_u kT}{2\pi\hbar^2} \right)^{3/2} \cdot \exp \left( -\frac{S_{1n}(N+1,Z)}{kT} \right), \label{eq:detailedbalance}
\end{align}
where $\expval{\sigma v}_{(n,\gamma)}$ is the velocity-integrated neutron capture cross section for a nucleus with $N$ neutrons and $Z$ protons ($A\equiv N+Z$), $G(N,Z)$ is the partition function for the nucleus $(N,Z)$ ($G(1,0)$ is the partition function for neutron), $m_u$ is the mass of a nucleon, and $T$ is the temperature of the environment. 
\textcolor{black}{
This shows that in addition to astrophysical conditions such as temperature $T$, the uncertainty in the reverse reaction rate, consequently the uncertainty of the final abundance pattern, arises from the nuclear physics inputs such as the integrated neutron capture cross section $\expval{\sigma v}_{(n,\gamma)}$, the partition functions $G(N,Z)$, and one-neutron separation energies $S_{1n}$.
}
\textcolor{black}{
Note that the calculations of other reaction and decay rates are also dependent on the nuclear structure and masses.
}

The mass models included in our ensemble are the Duflo-Zucker mass model with 29 parameters (DZ29) \cite{Duflo1995}, FRDM2012 \cite{Moller2016}, HFB31 \cite{Goriely2016}, KUTY05 \cite{Koura2005}, ETFSI2 \cite{Aboussir1995}, and WS4 \cite{Wang2014}.
\textcolor{black}{These models were chosen based on their popularity in $r$-process studies as well as the public availability of the data. When multiple versions of the same type of mass model are available, a newer version was selected. The included mass models are largely phenomenological in nature, and little has been investigated regarding uncertainty quantification.}

In some of our numerical experiments, we take the $S_{1n}$ values from the AME2020~\cite{Wang2021} as experimental data. 
When evaluating the quality of uncertainty estimates for unseen data, we use the $S_{1n}$ values from the AME2003~\cite{Audi2003} for constructing our models (we refer to them as ``training data'') and then test them with the new data in the AME2020. 
In the AME2020, 318 new $S_{1n}$ measurements with $Z=16$--105 are available compared to the AME2003. The new data points in the AME2020 compared to the AME2003 are shown in Figure~\ref{fig:Sn-AME2003_AME2020}.

We consider four different ways to categorize the $S_{1n}$ data. The first category is the data for the whole chart of nuclides, which employs all the available experimental data in \textcolor{black}{$Z=16$--105} at once. The second and third are data for each isotopic and isotonic chain, respectively. This focuses on the evolution of the $S_{1n}$ values as a function of proton and neutron number (isotopic and isotonic, respectively). The other category is isobaric (equal mass number $A$), which is relevant to the trend of $\beta$-decay $Q$-values. 

\textcolor{black}{
The reason for considering EBMA models for each isotopic, isotonic, or isobaric chain is that in many experimental and theoretical studies, the trend of the data or theoretical predictions in a specific chain is often of interest, and the EBMA model constructed with the data for the entire chart of nuclides may not necessarily capture such a local trend. It would also allow assessment of how well each mass model performs in different regions of the chart of nuclides, and by reviewing the theoretical descriptions of the best performing models, i.e. models with largest weights, it may potentially provide insight into what is required for the more precise modeling of nuclear masses. However, we note that optimizing the parameters to capture the local trend may cause overfitting. Especially, when EBMA models are constructed for each chain across the chart of nuclides, the total number of parameters is significantly more than in the case where a single EBMA model is constructed using the $S_{1n}$ values for the whole chart of nuclides. Therefore, a careful assessment of the quality of the uncertainty estimate is necessary. We will also investigate this aspect in Section~\ref{sec:ResultsDiscussion}.}

\textcolor{black}{
Except for Section~\ref{subsec:biascorrected} where the effect of bias correction is investigated, EBMA models are constructed without bias correction (denoted as ``raw'') for the $S_{1n}$ values predicted by each mass model.
}


\section{\label{sec:ResultsDiscussion}Results and Discussion}


\subsection{Comparison with experimental data}
\textcolor{black}{
To investigate the performance of the EBMA model in reconstructing the experimental $S_{1n}$ values, an EBMA model was constructed using the $S_{1n}$ data of all the nuclei with $16 \leq Z \leq 105$ from the AME2020 (referred to as the ``whole chart EBMA model'').
}
\textcolor{black}{
The whole chart EBMA model was constructed without bias correction of the mass models (raw).
}

\textcolor{black}{
Table~\ref{tab:WholeWeight} lists the 95~\% posterior highest density intervals (HDIs), which are the narrowest intervals that include 95~\% of the posterior distributions, of the whole chart EBMA model. The posterior weight, which can be interpreted as the probability of the model being the best one, is the largest for the WS4 model, followed by FRDM2012 and DZ29. Further analysis of the weights is provided in Section~\ref{subsec:weights}.}

The nominal predictions of the EBMA model are taken as the mode of the posterior predictive distributions of $S_{1n}$. The posterior distributions of weights and standard deviations $\sigma_k$ (Equation~\ref{eq:NormalPredictive}) are determined from the AME2020 data through Bayesian inference. Since the AME2020 values are used both for fitting and evaluation of the performance, this analysis reveals how well the EBMA method can reproduce known experimental data using the constituent mass models. 

\begin{table}[b]
\caption{\label{tab:WholeWeight}%
95~\% posterior highest density intervals (HDI) of the EBMA weights and standard deviations (variances), fitted with all the AME2020 $S_{1n}$ data in $16\leq Z \leq 105$ (referred to as ``whole chart'' in the text). The notation $(a, b)$ denotes an interval with $a$ being the lower bound and $b$ being the upper bound, respectively. $\sigma_\mathrm{RMS}$ (defined in Equation~\ref{eq:sigmarms}) shows the root mean square error of each mass model with respect to the same AME2020 data. \textcolor{black}{Bias correction of mass models was not performed (raw).}
}
\begin{ruledtabular}
\begin{tabular}{c|ccc}
\makecell{{Mass model}\\{}\textcolor{black}{(raw)}}&
\textrm{Weight}&
\makecell{Standard\\{}deviation}&
\textrm{$\sigma_\text{RMS}$ [MeV]}\\
\colrule
WS4~\cite{Wang2014} & (0.459, 0.596) & (0.183, 0.214) & 0.239\\
FRDM12~\cite{Moller2016} & (0.143, 0.251) & (0.129, 0.174) & 0.312\\
DZ29~\cite{Duflo1995} & (0.113, 0.229) & (0.125, 0.245) & 0.271\\
KUTY05~\cite{Koura2005} & (0.034, 0.130) & (0.140, 0.322) & 0.753\\
HFB31~\cite{Goriely2016} & (0.000, 0.027) & (0.183, 0.214) & 0.428\\
ETFSI2~\cite{Aboussir1995} & (0.000, 0.027) & (0.026, 0.761) & 0.828\\
\end{tabular}
\end{ruledtabular}
\end{table}

\begin{figure}[ht]
\includegraphics[width=\linewidth]{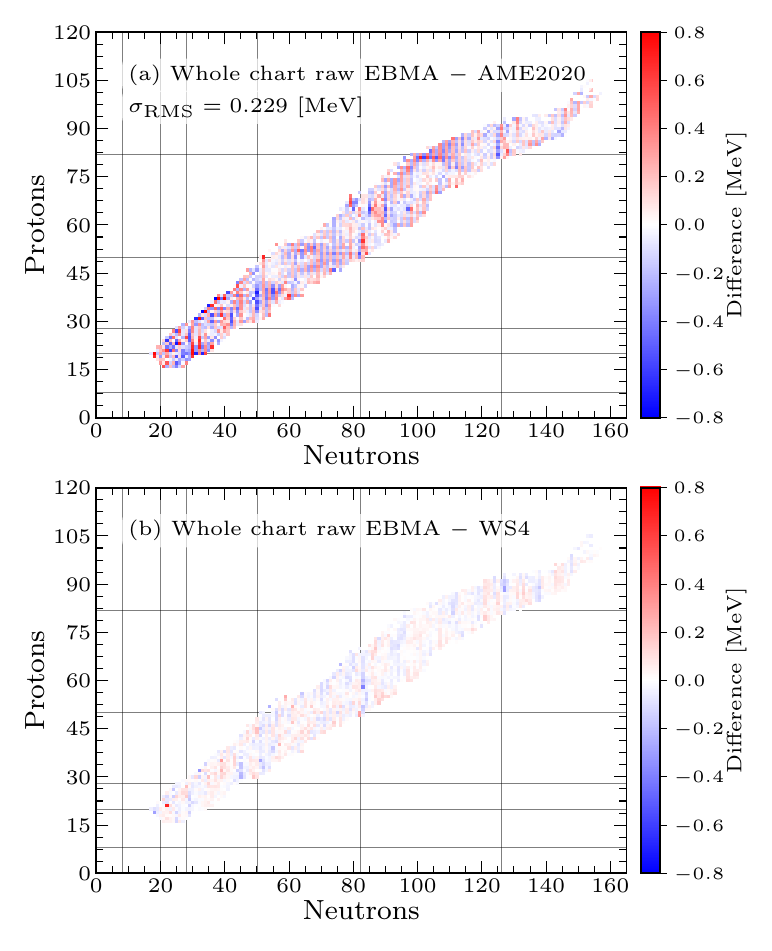}
\caption{\label{fig:Sn-rms}
\textcolor{black}{
Deviation and root mean square (RMS) error $\sigma_\text{RMS}$ [MeV] of the neutron separation energies $S_{1n}$ reconstructed by the EBMA model whose parameters are optimized using the whole AME2020 data (Whole chart raw EBMA), compared to the AME2020 data themselves (a), and the differeces of the predictions of the EBMA model from the WS4 model, which is the model in the ensemble with the largest weight (b). 
``raw'' indicates that bias correction was not performed for the models in the ensemble.
}
}
\end{figure}

Figure~\ref{fig:Sn-rms} shows the deviations of the modes of the EBMA poterior predictive for $S_{1n}$ from the AME2020 values. 
The root mean square error ($\sigma_\text{RMS}$) shown in the figure is defined as
\begin{align}
    \sigma_\text{RMS} = \sqrt{ \frac{\sum_{(N,Z)} \left(S_{1n}^{\text{AME2020}}{(N,Z)}-{S_{1n}^{\text{Model}}}{(N,Z)}\right)^2}{N^\text{AME2020}}}, \label{eq:sigmarms}
\end{align}
where $(N,Z)$ represents pairs of neutron number $N$ and proton number $Z$ of nuclei in the AME2020 whose $S_{1n}$ values are used for the fit. $N^\text{AME}$ is the total number of such nuclei ($N^\text{AME2020}\equiv\sum_{(N,Z)}$). 
\textcolor{black}{
For the EBMA model shown in Figure~\ref{fig:Sn-rms}, which took into account the nuclei with $16 \leq Z \leq 105$, $N^\text{AME2020}=2021$.
}
$S_{1n}^{\text{AME2020}}(N,Z)$ and $S_{1n}^{\text{Model}}(N,Z)$ are the $S_{1n}$ values for a nucleus $(N,Z)$ from the AME2020 and the mode of the posterior predictive distribution of $S_{1n}$ given by a EBMA model (or the nominal prediction of a specific mass model in Table~\ref{tab:WholeWeight}), respectively.  

\textcolor{black}{
The value of $\sigma_\text{RMS}$ of the whole chart EBMA model (Figure~\ref{fig:Sn-rms} panel (a)) shows a slightly better $\sigma_\mathrm{RMS}$ (0.229~MeV) than the best performing model,}
which is the WS4 model \cite{Wang2014} with $\sigma_\mathrm{RMS}^\mathrm{WS4}=0.239$~MeV (see Table~\ref{tab:WholeWeight}).
\textcolor{black}{
As shown in the panel~(b) of Figure~\ref{fig:Sn-rms}, while the constructed EBMA model is largely based on the WS4 model, the contribution of the other mass models in the ensemble is present throughout the chart of nuclides.
}


\subsection{\label{subsec:weights}Parameters in the EBMA models}
\textcolor{black}{
As shown in Table~\ref{tab:WholeWeight}, the top three mass models in the size of weight corresponds to the models with the smallest $\sigma_\text{RMS}$ with respect to the AME2020. Interestingly, while $\sigma_\mathrm{RMS}$ of the HFB31 model is significantly smaller than KUTY05 or ETFSI2, the assigned weight is one of the smallest. This implies that, although HFB31 can reproduce the experimental $S_{1n}$ values relatively well, the addition of the mass model does not necessarily contribute to improving the overall fit to the experimental data.
}

\textcolor{black}{
Table~\ref{tab:WholeWeight} also shows that the standard deviations (variances) of the normal distributions in the mixture model do not correspond to the $\sigma_\text{RMS}$ values. This is because EBMA is a normal mixture model, that is, a weighted sum of normal distributions whose parameters are inferred simultaneously based on the experimental data, while $\sigma_\text{RMS}$ is calculated separately for each model.
}

To investigate the weights in EBMA, in addition to the whole chart EBMA model discussed above, EBMA models were further constructed for:
\begin{itemize}
    \item each isotopic chain,
    \item each isotonic chain, and
    \item each isobaric chain,
\end{itemize}
respectively, using the AME2020 $S_{1n}$ data with the raw mass models.


\begin{figure}
\includegraphics[width=0.95\linewidth]{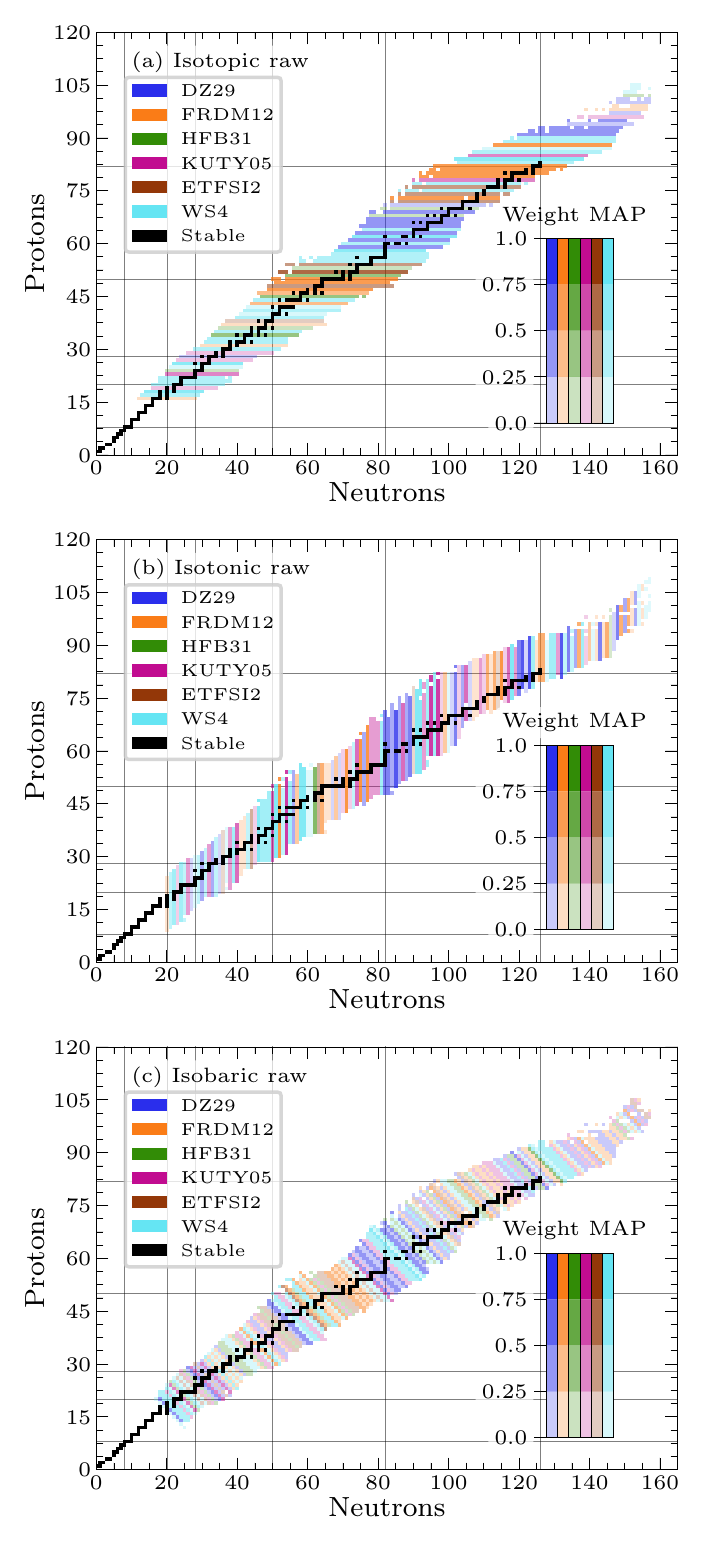}
\caption{\label{fig:Sn-weight}Maximum a posteriori (MAP) values of the largest weight in the EBMA ensemble determined from the AME2020 data for each (a) isotopic chain, (b) isotonic chain, and (c) isobaric chain. ``raw'' indicates that bias correction was not performed for the models in the ensemble.}
\end{figure}

The colors in Figure~\ref{fig:Sn-weight} show the mass model with the largest MAP value (maximum a posteriori; mode of posterior distribution) of weight within the ensemble for each isotopic (panel (a)), isotonic (panel (b)), and isobaric chain (panel (c)). The color scale represents the value of the MAP value of the weight. 
Weight can be understood as the posterior probability of the mass model being the best \cite{Raftery2005}, based on the training using the AME2020 data. 

\textcolor{black}{
In Figure~\ref{fig:Sn-weight}, especially for the isotopic (panel (a)) and isotonic (panel (b)) EBMA models, clustering of the same models in specific regions is visible. For the isotopic EBMA models, it can be seen that the FRDM2012 model tends to have the largest weights just below the proton magic numbers $Z=50$ and 82. The presence of the DZ29 model is also notable in $59\leq Z\leq71$ for odd proton numbers and in $91 \leq 95$. The WS4 model has largest weights at various locations on the chart of nuclides. For the isotonic EBMA models, the most notable trend is the presence of the DZ29 model just above the neutron magic number $N=82$. It can also be seen that the KUTY05 model is much more present across the chart of nuclides compared to the isotopic case. On the other hand, such clustering of specific model is less visible in the isobaric case.
}

\textcolor{black}{
This analysis shows that, by inferring the weights, it is possible to quantify the performance of each mass model in the ensemble in different regions of the chart of nuclides. Some mass models have been shown to perform better than others in reproducing experimental $S_{1n}$ values in specific regions. By further analyzing the well performing models, it may be possible to gain insight into what theoretical description of the nuclei can effectively reliably reproduce and predict nuclear masses in different parts of the chart of nuclides.
}


\subsection{\label{subsec:uncertainty}Uncertainty quantification with EBMA}
One of the main goals of this study is to quantify the uncertainty of theoretical $S_{1n}$ values when a variety of mass models are available. EBMA estimates it by creating a weighted average of a collection of mass models based on the performance of each model during the training. In the EBMA model, predictive uncertainty includes not only the spread of the forecasts among the members of the ensemble, but also takes into account the weighted variance of each member model according to the performance during the training \cite{Raftery2005}. 

\subsubsection{Size of uncertainty estimates}

\begin{figure*}[ht]
\includegraphics[width=\linewidth]{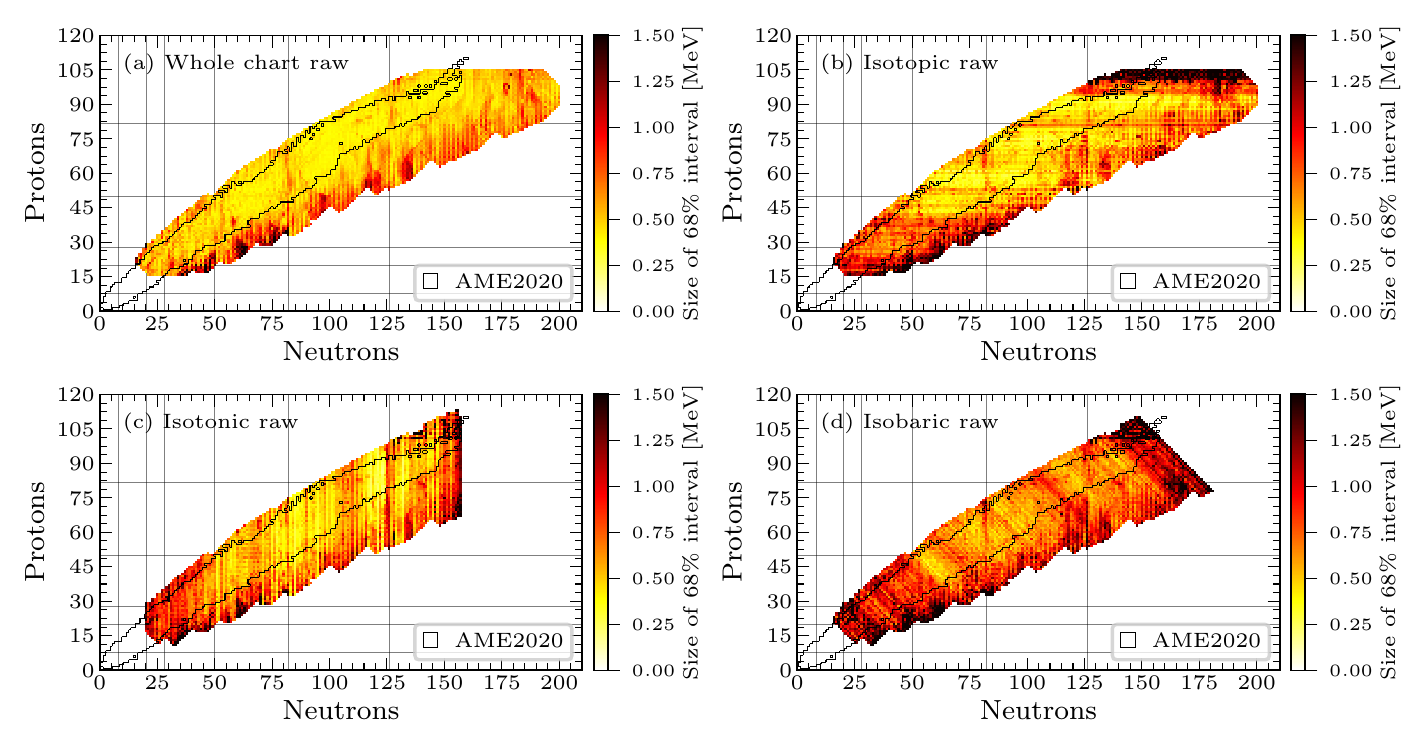}
\caption{\label{fig:Sn-hdi68}The sizes of 68\% highest density intervals (HDIs) of the EBMA models across the chart of nuclides, fitted with the AME2020 $S_{1n}$ data for (a) the whole chart of nuclides, (b) each isotopic chain, (c) each isotonic chain, and (d) each isobaric chain. 
The area enclosed by the black contour shows where the AME2020 data exist.
The charts with isotonic and isobaric fits are truncated at large neutron number and mass number, respectively, because there are not enough data points within the chains to determine the EBMA parameters.
``raw'' indicates that bias correction was not performed for the models in the ensemble.
}
\end{figure*}

\begin{figure*}
\centering
\includegraphics[width=1\linewidth]{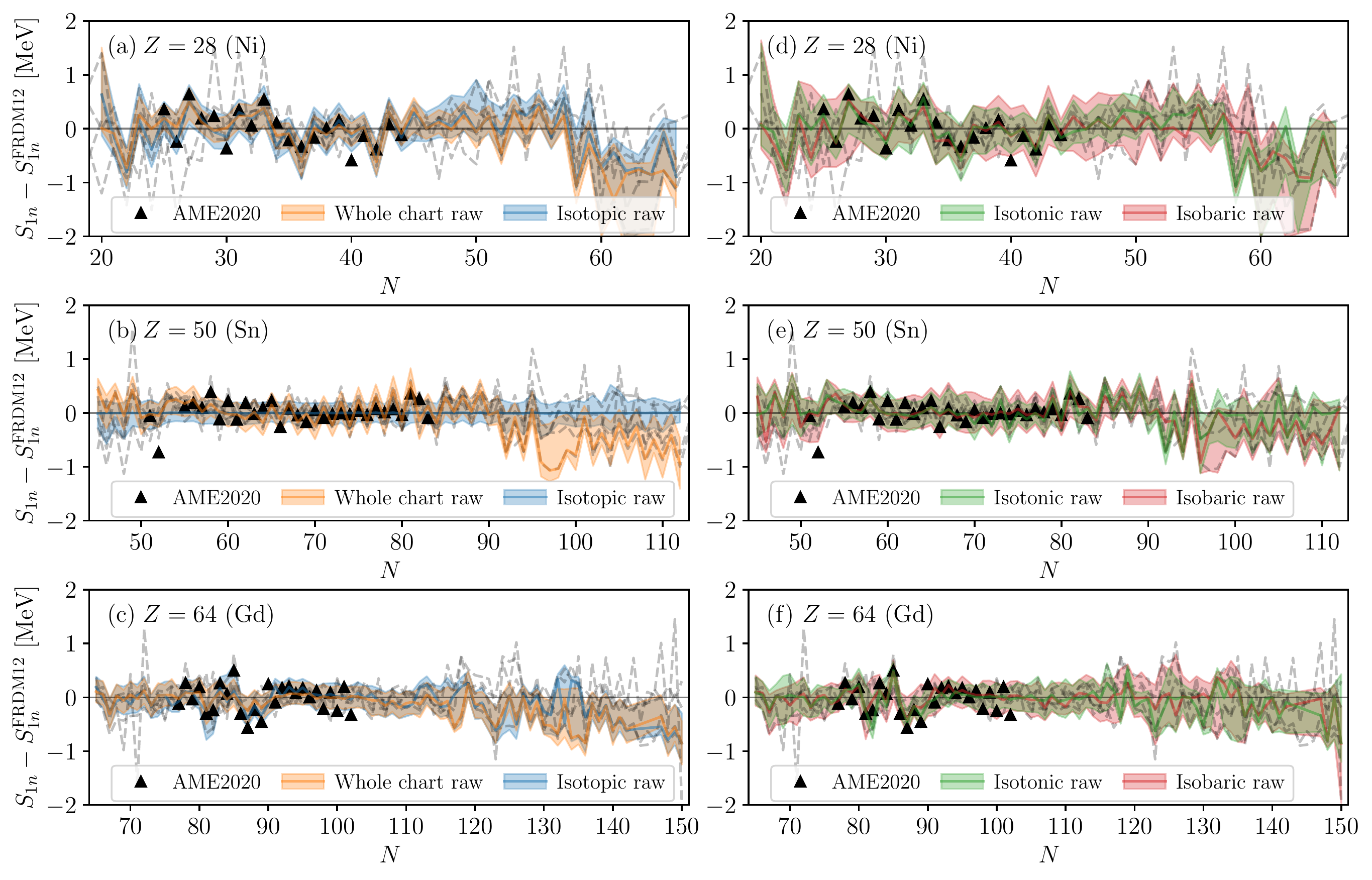}
\caption{\label{fig:Sn-hdi68-isotopes}
Trends of the sizes of 68\% highest density intervals of the EBMA models compared to the predictions by the FRDM2012 model for the $S_{1n}$ values in the $Z=28$ (Ni) (panels (a) and (d)), $Z=50$ (Sn) (panels (b) and (e)), and $Z=64$ (Gd) (panels (c) and (f)) isotopic chains, similarly to Figure~\ref{fig:Sn-models-comparison}. Panels (a--c) show the EBMA models fitted with the AME2020 $S_{1n}$ data for the whole chart of nuclides (orange) and each isotopic chain (blue), and (d--f) show the models for each isotonic chain (green) and each isobaric chain (red). The solid lines associated with the bands show the modes of the predictive distributions. The gray dashed lines are the mass models shown in Figure~\ref{fig:Sn-models-comparison}.
``raw'' indicates that bias correction was not performed for the models in the ensemble.
}
\end{figure*}

Figure~\ref{fig:Sn-hdi68} shows the size of the 68\% highest density interval (HDI), which is roughly analogous to the $\pm 1\sigma$ interval of the normal distribution. EBMA models were fitted for the whole chart of nuclides (panel (a)), each isotopic chain (panel (b)), each isotonic chain (panel (c)), and each isobaric chain (panel (d)), respectively.
\textcolor{black}{
Bias correction was not performed for the mass models in the ensemble (raw).}
\textcolor{black}{Figure~\ref{fig:Sn-hdi68-isotopes} also shows the 68\% HDIs, focusing on the $Z=28$ (nickel), $Z=50$ (tin), and $Z=64$ (gadolinium) isotopes, compared to the mass models (gray dashed lines) shown in Figure~\ref{fig:Sn-models-comparison}, relative to the nominal predictions of the FRDM2012 model.
The panels (a)--(c) show the EBMA models fitted for the whole chart of nuclides and each isotopic chain and (d)--(f) show the models for each isotopic chain and each isobaric chain for the same set of nuclei.}
The fits are performed using the AME2020 data, and predictions are made for all the nuclei available in all the member mass models within the ensemble. 
In all cases, it can be seen that the size of the uncertainty is more constrained where the data exist (black contour in Figure~\ref{fig:Sn-hdi68} \textcolor{black}{and black triangles in Figure~\ref{fig:Sn-hdi68-isotopes}}), but increases towards the edge of the chart of nuclides. This is especially visible in the neutron-rich direction, 
\textcolor{black}{
where many of the nuclear masses are yet to be experimentally measured. 
}
This is a reflection of the fact that the predictions of the mass models that make up the ensemble start to diverge as we move further away from the last data point, as shown in Figure~\ref{fig:Sn-models-comparison}.

Comparing the four plots in Figure~\ref{fig:Sn-hdi68}, the increase in the size of uncertainty in the neutron-rich region is the smallest for the fit using the whole chart of nuclides (panel (a)). This is because the weights for the whole chart EBMA model are determined using all available data, whereas for each isotopic, isotonic, and isobaric chain, the weights are determined only from the data in each chain. 
\textcolor{black}{
The smaller number of training data points (experimental $S_{1n}$ values) in each chain also results in larger uncertainties for reproducing the experimental values, compared to the fit with the whole chart data.
}

\textcolor{black}{
As shown in Figure~\ref{fig:Sn-hdi68-isotopes}, for some isotopic chains, the uncertainty estimates are different between different types of fits. Especially, for $Z=50$ isotopes, it can be seen in panel (b) that the uncertainty estimated by the whole chart EBMA model constantly falls below 0 (the predictions by FRDM2012) beyond $N=90$, while the isotopic EBMA model is dominated by FRDM2012. It is not yet possible to experimentally test the predictions for such neutron-rich nuclei. However, differences in the predictions in neutron-rich regions may have a significant impact on the prediction of observables from the $r$-process such as the abundance pattern. 
}




\subsubsection{\label{subsubsec:Quality}Quality of uncertainty estimates}
We now investigate the quality of the uncertainty estimate. For this purpose, we construct EBMA models and quantify the prediction uncertainties using the $S_{1n}$ data for nuclei with $16\leq Z \leq 105$ from the AME2003 \cite{Audi2003} (we will refer to the data as ``training data''), then evaluate the quality of the uncertainties based on the new $S_{1n}$ data in the AME2020 \cite{Wang2021}. In the AME2020, there are 318 $S_{1n}$ values newly registered since the AME2003. Bias correction was not performed for the mass models in the ensemble (raw).

\begin{figure*}
\includegraphics[width=\linewidth]{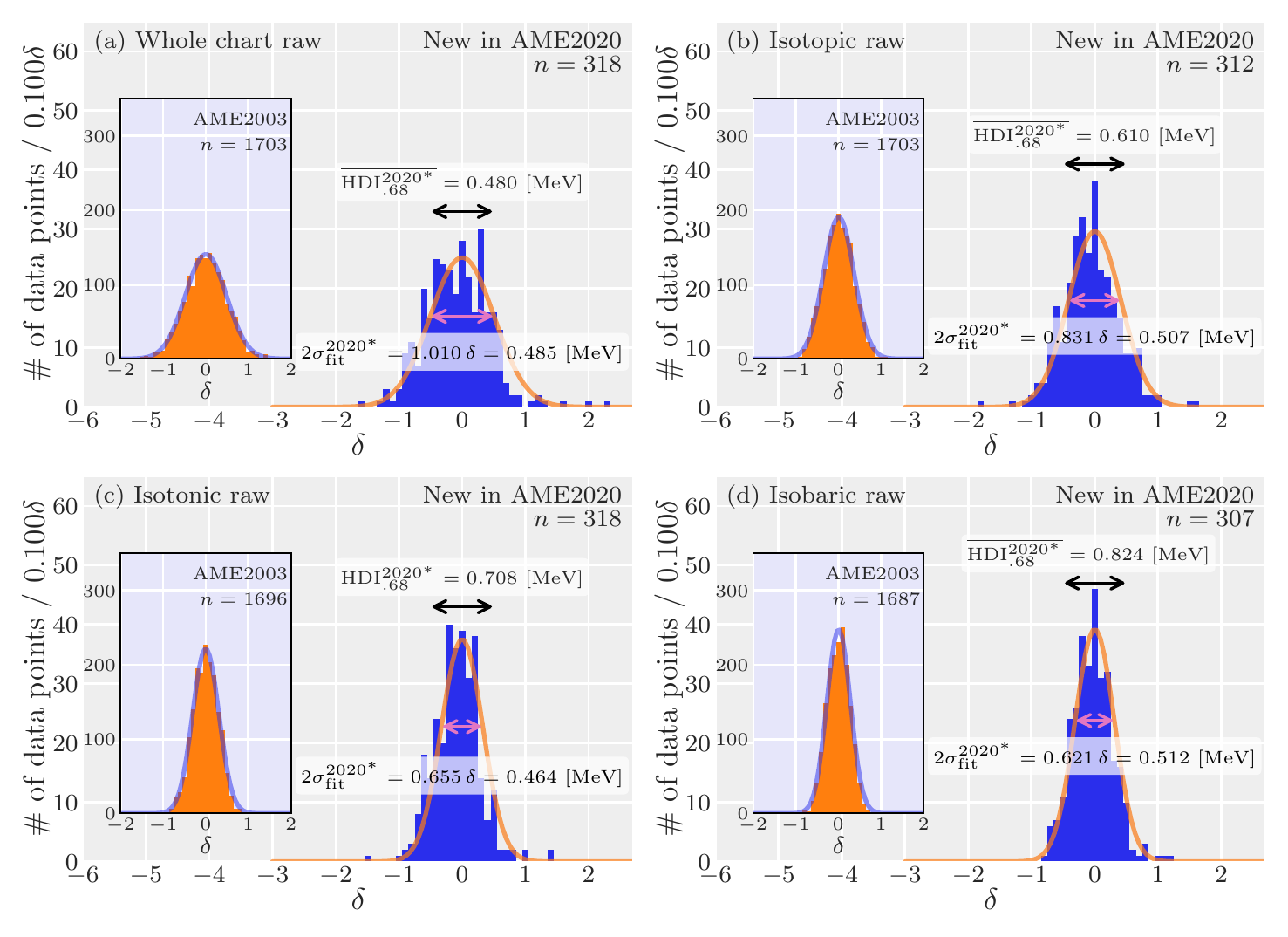}
\caption{\label{fig:Sn-dist-hdi68}Distributions of the new data points in the AME2020 compared to the AME2003, with respect to the 68\% HDIs predicted by the EBMA models fitted with the AME2003 data for (a) the whole chart of nuclides, (b) each isotopic chain, (c) each isotonic chain, and (d) each isobaric chain. The definition of $\delta$ is given in Equation~\ref{eq:delta}. ``raw'' indicates that bias correction was not performed for the models in the ensemble.}
\end{figure*}

Figure~\ref{fig:Sn-dist-hdi68} shows the distribution of the new $S_{1n}$ data relative to the sizes of uncertainties given by the EBMA models fitted with the data for the whole chart of nuclides (panel (a)), each isotopic chain (panel (b)), isotonic chain (panel (c)), and isobaric chain (panel (d)). Note that the number of new data points included in the fit ($n$ in Figure~\ref{fig:Sn-dist-hdi68}) is not necessarily 318 for some cases. This is because at the edge of the chart of nuclides, there are fewer experimental data points and the posterior weights do not converge for some chains. Such chains were excluded from the fits.
\textcolor{black}{
The summary of the analysis is shown in the upper half of Table~\ref{tab:68summary}. In the following, we describe our analysis and findings.
}

\textcolor{black}{
The size of the estimated uncertainty is represented by the 68\% highest density interval (HDI$_{.68}$), which is the narrowest interval that contains 68\% of the distribution, of the posterior predictive distribution of the $S_{1n}$ values.
} 
To investigate how the new experimental $S_{1n}$ values are distributed relative to the predicted HDI$_{.68}$, we define $\delta$, which represents an experimental $S_{1n}$ value normalized by the size of HDI$_{.68}$. Let $h^ {\mathrm{low}}$ and $h^\mathrm{up}$ represent the lower and upper boundaries of the HDI$_{.68}$, respectively, then
\begin{equation}
    \delta = \frac{S_{1n} - h^\mathrm{low}}{h^\mathrm{up}-h^\mathrm{low}}-0.5 \label{eq:delta},
\end{equation}
where 0.5 is subtracted to symmetrize the distribution around 0. This means that \textcolor{black}{$\delta=-0.5$ and $0.5$} correspond to the $S_{1n}$ values at the lower and upper boundaries of the HDI$_{.68}$, respectively.

\textcolor{black}{
$\overline{\mathrm{HDI}_{.68}^{2020^*}}$ in Figure~\ref{fig:Sn-dist-hdi68} and Table~\ref{tab:68summary} shows the average size of the 68\% intervals for the new data in the AME2020 (denoted as $2020^*$). On average, the whole chart EBMA model provides the most constrained size of the uncertainty of 0.480~MeV. Table~\ref{tab:68summary} also shows the root mean square errors of the modes of the posterior predictive distributions of $S_{1n}$ given by the EBMA models from the experimental $S_{1n}$ values from the AME2003 ($\sigma_\mathrm{RMS}^{2003}$), similarly to Eq.~\ref{eq:sigmarms}. While the models fitted for each isotopic/isotonic/isobaric chain exhibit smaller $\sigma_\mathrm{RMS}^{2003}$ values, the sizes of $\overline{\mathrm{HDI}_{.68}^{2020^*}}$ are larger than the whole chart EBMA model. This is because the EBMA models are constructed and fine-tuned for each chain, which also means that the total number of parameters across the chart is significantly larger than the fitting of the whole chart with a single EBMA model. At the same time, the parameters for each chain are inferred from a smaller number of data points compared to the whole chart EBMA model; therefore, the parameters are less sharply determined. This suggests that the large total number of parameters of the EBMA models constructed for individual chains across the chart of nuclides may lead to overfitting. However, also considering the associated uncertainty, the overfitting may partially be mitigated by the increased uncertainty through the Bayesian framework. We will further investigate this issue below.
}

\begin{table*}[!htbp]
\caption{\label{tab:68summary}
Quantities that characterize the uncertainty estimates and their quality for different EBMA models, which are fitted to the ${S_{1n}}$ data for nuclei with $16\leq Z \leq 105$ from the AME2003. $\sigma_\mathrm{RMS}$ is the root mean square error calculated similarly to Eq.~\ref{eq:sigmarms}, $\overline{\mathrm{HDI}_{.68}}$ is the average size of the 68\% highest density intervals, $2\sigma_{\mathrm{fit}}$ is the spread of the experimental $S_{1n}$ values around the center of $\mathrm{HDI}_{.68}$ approximated by a normal distribution, and the $\mathrm{HDI}_{.68}$ coverage is the ratio of experimental $S_{1n}$ values that fall within $\mathrm{HDI}_{.68}$. The superscripts ``2003'' and ``$2020^*$'' denote the AME2003 data and the new data in the AME2020 (absent in the AME2003), respectively. The upper half shows the results without bias correction (raw) and the lower half with bias correction (corrected).%
}
\begin{ruledtabular}
\begin{tabular}{cccccccc}
\makecell{EBMA model}&
\makecell[l]{$\sigma^{2003}_\text{RMS}$\\{}[MeV] }&
\makecell{$\overline{\mathrm{HDI}^{2003}_{.68}}$\\{}[MeV]} &
\makecell{$2\sigma^{2003}_{\mathrm{fit}}$\\{}[MeV]} &
\makecell{$\mathrm{HDI}_{.68}^{2003}$\\{}coverage [\%]} &
\makecell{$\overline{\mathrm{HDI}^{2020^*}_{.68}}$\\{}[MeV]} &
\makecell{$2\sigma^{2020^*}_{\mathrm{fit}}$\\{}[MeV]} &
\makecell{$\mathrm{HDI}^{2020^*}_{.68}$\\{}coverage [\%]} \\
\colrule
Whole chart (raw) &  0.239 & 0.466 & 0.448 & 71.5 & 0.480 & 0.485 & 69.5 \\
Isotopic (raw) &  0.218 & 0.562 & 0.401 & 85.1 & 0.610 & 0.507 & 75.3 \\
Isotonic (raw) & 0.201 & 0.598 & 0.368 & 92.5 & 0.708 & 0.464 & 82.7 \\
Isobaric (raw) &  0.219 & 0.746 & 0.411 & 95.6 & 0.824 & 0.512 & 89.6 \\
\colrule
Whole chart (corrected) &  0.238 & 0.466 & 0.449 & 71.3 & 0.480 & 0.490 & 70.1 \\
Isotopic (corrected) &  0.206 & 0.517 & 0.373 & 84.4 & 0.706 & 0.664 & 72.4 \\
Isotonic (corrected) & 0.134 & 0.370 & 0.219 & 89.8 & 0.658 & 0.627 & 68.3 \\
Isobaric (corected) &  0.170 & 0.618 & 0.310 & 95.4 & 0.804 & 0.614 & 82.1 \\
\end{tabular}
\end{ruledtabular}
\end{table*}

\textcolor{black}{
Insets of Figure~\ref{fig:Sn-dist-hdi68} show that the distribution of the training $S_{1n}$ values from the AME2003 around the center of $\mathrm{HDI}_{.68}$ approximately follows a normal distribution. Performing a similar fit to the new data in the AME2020, we obtain the approximate 68\% interval of the distribution of the new experimental $S_{1n}$ values in the AME2020 around the center of $\mathrm{HDI}_{.68}$, which is denoted as $2\sigma^{2020^*}_{\mathrm{fit}}$. Comparing the sizes of $\overline{\mathrm{HDI}_{.68}^{2020^*}}$ and $2\sigma^{2020^*}_{\mathrm{fit}}$ for the whole chart EBMA model, it can be seen that the sizes are comparable, which shows that this EBMA model can appropriately estimate the uncertainty for the new data. On the other hand, the sizes of $\overline{\mathrm{HDI}_{.68}^{2020^*}}$ for the other EBMA models are larger than $2\sigma^{2020^*}_{\mathrm{fit}}$, indicating that the uncertainty estimates are inflated (underconfident). The tendency of inflated uncertainty for the EBMA with individual chains can also be seen from the comparison of the corresponding quantities for the AME2003 training data ($\overline{\mathrm{HDI}_{.68}^{2003}}$ and $2\sigma^{2003}_{\mathrm{fit}}$), shown in Table~\ref{tab:68summary}.
}

\textcolor{black}{
This observation is further supported by looking at the ratios of the $S_{1n}$ values that fall within $\mathrm{HDI}_{.68}$ for the AME2003 and the new data in the AME2020, also shown in Table~\ref{tab:68summary} as the $\mathrm{HDI}^{2003}_{.68}$ coverage and the $\mathrm{HDI}^{2020^*}_{.68}$ coverage, respectively. Except for the whole chart EBMA model, the coverages significantly exceed 68\% for both data sets. Comparing the coverages of $\mathrm{HDI}^{2003}_{.68}$ and $\mathrm{HDI}^{2020^*}_{.68}$, except for the whole chart EBMA, the coverages of $\mathrm{HDI}^{2020^*}_{.68}$ are smaller than $\mathrm{HDI}^{2003}_{.68}$. This points to the fact that the quality of uncertainty estimates does not extrapolate well for the EBMA models constructed for individual chains. In other words, this can be seen as a sympton of overfitting to the training data. 
}

\textcolor{black}{
From this analysis, it can be concluded that the uncertainty estimates of the whole chart EBMA have desirable properties and extrapolate well to the new data in the AME2020. On the other hand, the EBMA models constructed for each isotopic/isotonic/isobaric chain tend to overfit the AME2003 data and at the same time provide underconfident uncertainty estimates. However, it is intriguing that the isotonic EBMA model resulted in the smallest values of $\sigma_\mathrm{RMS}^{2003}$, $\sigma_\mathrm{fit}^{2003}$, and $\sigma_\mathrm{fit}^{2020^*}$. This suggests the possibility that focusing on the isotonic chain may potentially provide a reliable way to extrapolate the $S_{1n}$ values if overfitting can be avoided and the size of the uncertainty estimate can be appropriately controlled. We note that the new data in the AME2020 lie just outside the AME2003 data, as shown in Fig.~\ref{fig:Sn-AME2003_AME2020}. Therefore, these results may not hold in a more neutron-rich region. Further development of the EBMA method will be carried out to put more emphasis on the neutron-rich nuclei. Without abundant data in the neutron-rich region, alternative constraints may also be necessary, e.g., the $r$-process observables, which are known to involve neutron-rich nuclei.
}

\textcolor{black}{
As discussed above, EBMA, especially the whole chart EBMA model, provides reasonable uncertainty estimates for the current data sets. Therefore, if a new experimental $S_{1n}$ value significantly deviates from the uncertainty estimates, it suggests that the value is unexpected (an outlier) based on the mass models in the ensemble and the training data. The current whole chart EBMA model predicts that the $S_{1n}$ values of \ce{^{66}_{33}As} ($N=Z=33$), \ce{^{70}_{35}Br} ($N=Z=35$), \ce{^{74}_{37}Rb} ($N=Z=37$), and \ce{^{102}_{50}Sn} ($N=52$ and $Z=50$) lie in the region where $\abs{\delta}>1.5$. Note that the edges of $\mathrm{HDI}_{.68}$ correspond to $\abs{\delta}=0.5$. The $S_{1n}$ values of \ce{^{66}_{33}As} ($N=Z=33$) and \ce{^{74}_{37}Rb} are also in $\abs{\delta}>1.5$ based on the isotopic EBMA model, and \ce{^{102}_{50}Sn} is in $\abs{\delta}>1.5$ based on the isotopic and isotonic EBMA models. The isobaric EBMA model did not detect any $S_{1n}$ values in $\abs{\delta}>1.5$. All the nuclei listed above are on or close to the $N=Z$ line. The nuclei with $N=Z$ tend to be more strongly bound than the neighboring nuclei (Wigner effect) and often the correction has to be explicitly incorporated into the mass models \cite{Lunney2003}. Therefore, the above results imply an imperfection in incorporating the Wigner effect. Thus, uncertainty quantification with EBMA may possibly point to a deficiency in the theoretical description of nuclei.
}

\subsection{\label{subsec:biascorrected}\textcolor{black}{Bias corrected EBMA models}}
\textcolor{black}{
As discussed in Section~\ref{subsec:EBMA}, Ref.~\cite{Raftery2005} suggests a linear bias correction of each model, prior to inference of the EBMA weights and standard deviations. This means determining the bias correction coefficients $a_k$ and $b_k$ for each model, where $a_k$ is the intercept (offset) and $b_k$ is the slope. $a_k + b_k m_k$ is then the prediction of the bias-corrected model based on the mass model $k$. In our case, $m_k$ is a vector of theoretical $S_{1n}$ values for the whole chart of nuclides or a specific isotopic/isotonic/isobaric chain, predicted by the model $k$.
}

\textcolor{black}{
The simplest approach to determine $a_k$ and $b_k$ is to perform linear regression, as shown in Ref.~\cite{Raftery2005}.
Another way to determine these parameters $a_k$ and $b_k$ is by Bayesian linear regression and taking the \emph{maximum a posteriori} (MAP) values, which would be a slightly more probabilistic treatment. In our case, the two approaches yield virtually identical values. 
The coefficients determined for each mass model using the entire AME2003 data are shown in Table~\ref{tab:bias}, for example. It can be seen that for some mass models, the coefficients $a_k$ deviate from $0$. 
}

\begin{table}[b]
\caption{\label{tab:bias}%
\textcolor{black}{
Bias-correction coefficients determined for different mass models from the MAP values of Bayesian linear regression (labeled as ``BLR MAP'') and usual linear regression (labeled as ``LR''), respectively, using all of the available (whole chart) one-neutron separation energy ($S_{1n}$) data from the AME2003~\cite{Audi2003}. $a_k$ and $b_k$ are the coefficients for offset (intercept) and slope, respectively, for mass model $k$. The values are rounded to two digits.
}
}
\begin{ruledtabular}
\begin{tabular}{c rrrr}
 & \multicolumn{2}{c}{$a_k$ } & \multicolumn{2}{c}{$b_k$}\\
\cmidrule(lr){2-3} \cmidrule(lr){4-5}
Mass model & BLR MAP   & LR        & BLR MAP   & LR \\
\midrule
WS4     & -0.01   & -0.01   & 1.00    & 1.00 \\
FRDM12  & 0.12    & 0.12    & 0.99    & 0.99 \\
DZ29    & -0.10   & -0.10   & 1.01    & 1.01 \\
KUTY05  & -0.06   & -0.06   & 1.01    & 1.01 \\
ETFSI2  & 0.38    & 0.38    & 0.95    & 0.95\\
HFB31   & 0.30    & 0.30    & 0.96    & 0.96\\
\end{tabular}
\end{ruledtabular}
\end{table}

\textcolor{black}{
The summary of the models that quantify the quality of the uncertainty estimates is shown in the lower half of Table~\ref{tab:68summary}. As expected with further calibration (bias correction) of the mass models in the ensemble, the values of $\sigma_\mathrm{RMS}^{2003}$ decreased compared to the raw EBMA models. For the bias-corrected whole chart EBMA model, since the corrections made by the coefficients for the dominating mass models (WS4, FRDM2012, and DZ29) are small, as shown in Table~\ref{tab:bias}, the difference from the raw model is smaller compared to other EBMA models. 
For the bias-corrected EBMA models for individual chains, indications of inflated uncertainty estimates are still present, following the same argument given for the raw EBMA models in the previous section.
}

\textcolor{black}{
A notable difference from the raw EBMA models is the more significant decrease in $\mathrm{HDI}^{2020^*}_{.68}$ coverage from the $\mathrm{HDI}^{2003}_{.68}$ coverage for the EBMA models for individual chains, indicating further overfitting problems.
}

\textcolor{black}{
In conclusion, since the linear bias correction further facilitates overfitting in the EBMA models for individual chains, which was already not as reliable as the whole chart EBMA model as discussed in Section~\ref{subsec:uncertainty}, and does not have a significant effect on the whole chart EBMA model, the use of linear bias correction would not be recommended.
}

\subsection{\label{subsec:modelselection}\textcolor{black}{Effect of the choice of mass models in the ensemble}}
\textcolor{black}{
So far we have only considered the ensemble consisting of the mass models listed in Table~\ref{tab:WholeWeight}, which are some of the most popular models in the $r$-process studies. Nevertheless, it is possible to consider ensembles of any combination of mass models, and it is important to understand how the choice of models in the ensemble affects the uncertainty estimates. 
}

\textcolor{black}{
In the case where an ensemble of mass models includes models with poor performance in reproducing the experimental values, in general one can expect that such models will be assigned small weights, as seen for the weight of ETFSI2 shown in Table~\ref{tab:68summary}, for example. This is because the likelihood (Eq.~\ref{eq:EBMALikelihood}) evaluates how well each mass model reproduces the experimental data. Models with small weights have a small effect on quantified uncertainty, as models in the ensemble have contributions to the uncertainty proportional to the weights, as shown in the expression of predictive variance (Eq.~\ref{eq:predvar}).
}

\textcolor{black}{As for the removal of models with large weights, the effect is not trivial. Since the WS4 model has the largest contribution to the whole chart EBMA model we have considered so far, we will investigate how the EBMA model is affected by removing the WS4 model from the ensemble. From this we aim to gain insight into the effect of the choice of mass models. Based on the quality of uncertainty estimates investigated in Sections~\ref{subsubsec:Quality} and \ref{subsec:biascorrected}, we limit ourselves to the whole chart EBMA model without bias correction. 
}

\textcolor{black}{
In order to compare with the inferred EBMA parameters shown in Table~\ref{tab:WholeWeight}, a new whole chart EBMA model was constructed using the same AME2020 data in $16\leq Z \leq 105$, using an ensemble without the WS4 model. The 95\% HDIs for the posterior weights and standard deviations are shown in Table~\ref{tab:WholeWeight_noWS4}, in order of the size of the posterior weights. It can be seen that the assignment of the weights is somewhat more democratic, without the dominant WS4 model. It is also notable that the DZ29 model now has the largest weight, while it previously had the third largest weights in the ensemble with WS4. This implies that DZ29 and WS4 may have provided somewhat similar predictions of $S_{1n}$, while the performance of WS4 is consistently better with respect to the experimental data; therefore, WS4 had absorbed the possible contribution of DZ29.
}

\begin{table}[b]
\caption{\label{tab:WholeWeight_noWS4}%
95~\% posterior highest density intervals (HDI) of the EBMA weights and standard deviations (variances) without the WS4 model in the ensemble, fitted with all the AME2020 $S_{1n}$ data in $16\leq Z \leq 105$ (whole chart), similarly to Table~\ref{tab:WholeWeight}. The notation $(a, b)$ denotes an interval with $a$ being the lower bound and $b$ being the upper bound, respectively. 
$\sigma_\mathrm{RMS}$ (defined in Equation~\ref{eq:sigmarms}) shows the root mean square error of each mass model with respect to the same AME2020 data, identical to Table~\ref{tab:WholeWeight}. Bias correction of mass models was not performed (raw).
}
\begin{ruledtabular}
\begin{tabular}{c|ccc}
\makecell{\textrm{Mass model}\\{\textcolor{black}{(raw)}}}&
\textrm{Weight}&
\makecell{Standard\\{}deviation}&
\textrm{$\sigma_\text{RMS}$ [MeV]}\\
\colrule
DZ29 & (0.360,  0.484) & (0.207, 0.248) & 0.271 \\
FRDM12 & (0.273, 0.383) & (0.144, 0.175) & 0.312 \\
KUTY05 & (0.119, 0.230) & (0.158, 0.243) & 0.753 \\
HFB31 & (0.030, 0.089) & (0.128 0.264) & 0.428 \\
ETFSI2 & (0.002, 0.030) & (0.029, 0.663) & 0.828 \\
\end{tabular}
\end{ruledtabular}
\end{table}

\textcolor{black}{
Similarly to the Section~\ref{subsubsec:Quality}, we now construct the whole chart EBMA model with the AME2003 ${S_{1n}}$ data and test the quality of uncertainty estimates using the new data in the AME2020. The summary of performance is shown in Table~\ref{tab:68summary_noWS4}, again focusing on 68\% highest density intervals ($\mathrm{HDI}_{.68}$) of the predictive distributions of $S_{1n}$. In terms of reproducing the experimental values in the AME2003 (the training data), $\sigma^{2003}_\text{RMS}$ is 0.257~MeV, increasing from the 0.239~MeV with the ensemble with WS4. It is still an improvement over individual mass models, since the root mean square error of DZ29 is 0.274~MeV, which is the smallest in the current ensemble. Although the sizes of both $\overline{\mathrm{HDI}^{2003}_{.68}}$ and $2\sigma^{2003}_{\mathrm{fit}}$ are also about 0.3~MeV larger than those with the ensemble with WS4, the coverage of $\mathrm{HDI}_{.68}^{2003}$ is the same (71.5\%), showing that the quality of the uncertainty estimates for the training data is still adequate.
}

\textcolor{black}{
We now look at the quality of uncertainty estimates for the new data in the AME2020. A notable difference from the case with the ensemble with WS4 is that $2\sigma^{2020^*}_{\mathrm{fit}}$, the average size of the deviation from the center of $\mathrm{HDI}^{2020^*}_{.68}$, is significantly larger than the size of $\overline{\mathrm{HDI}^{2020^*}_{.68}}$, resulting in the coverage of $\mathrm{HDI}^{2020^*}_{.68}$ of 66.7\%. The decrease in coverage from the AME2003 data is also greater than in the case with the ensemble with WS4, suggesting that the performance of extrapolation in uncertainty estimates is slightly degraded. However, considering that the ideal coverage is 68\%, overall quality of the uncertainty estimates is still decent.
}

\begin{table*}[!htbp]
\caption{\label{tab:68summary_noWS4}
Similar table to Table~\ref{tab:68summary}, but for the whole chart EBMA model based on the ensemble without WS4. It shows the quantities that characterize the uncertainty estimates and their quality for different EBMA models, which are fitted to the ${S_{1n}}$ data for nuclei with $16\leq Z \leq 105$ from the AME2003. $\sigma_\mathrm{RMS}$ is the root mean square error calculated similarly to Eq.~\ref{eq:sigmarms}, $\overline{\mathrm{HDI}_{.68}}$ is the average size of the 68\% highest density intervals, $2\sigma_{\mathrm{fit}}$ is the spread of the experimental $S_{1n}$ values around the center of $\mathrm{HDI}_{.68}$ approximated by a normal distribution, and the $\mathrm{HDI}_{.68}$ coverage is the ratio of experimental $S_{1n}$ values that fall within $\mathrm{HDI}_{.68}$. The superscripts ``2003'' and ``$2020^*$'' denote the AME2003 data and the new data in the AME2020 (absent in the AME2003), respectively. 
}
\begin{ruledtabular}
\begin{tabular}{cccccccc}
\makecell{EBMA model (no WS4)}&
\makecell[l]{$\sigma^{2003}_\text{RMS}$\\{}[MeV] }&
\makecell{$\overline{\mathrm{HDI}^{2003}_{.68}}$\\{}[MeV]} &
\makecell{$2\sigma^{2003}_{\mathrm{fit}}$\\{}[MeV]} &
\makecell{$\mathrm{HDI}_{.68}^{2003}$\\{}coverage [\%]} &
\makecell{$\overline{\mathrm{HDI}^{2020^*}_{.68}}$\\{}[MeV]} &
\makecell{$2\sigma^{2020^*}_{\mathrm{fit}}$\\{}[MeV]} &
\makecell{$\mathrm{HDI}^{2020^*}_{.68}$\\{}coverage [\%]} \\
\midrule
Whole chart (raw) &  0.257 & 0.495 & 0.479 & 71.5 & 0.506 & 0.542 & 66.7 \\
\end{tabular}
\end{ruledtabular}
\end{table*}

\textcolor{black}{
We again note that the new data in the AME2020 lie just outside of the available data in the AME2003, therefore, the current results may not apply to the extrapolation to the more neutron-rich region. With this in mind, the current results show that, while the removal of the best-performing mass model WS4 slightly degrades the performance in extrapolating the uncertainty estimates of $S_{1n}$, the whole chart EBMA model still provides decent uncertainty estimates. This in turn suggests that, although a small root mean square error does not necessarily mean a large contribution to the ensemble, as shown from the weights of HFB31, including performant mass models would likely contribute to constraining the size of uncertainty and improving the quality in extrapolation of uncertainty estimates. It would also be recommended to include a wide range of models in the ensemble, even when some of the models are similar in nature, since it is difficult to know the possible contributions of each model. If there are models with similar but less performant predictions, their contributions will automatically be absorbed by the more performant models. 
}

\section{\label{sec:conclusion}Conclusions}
We have explored the possibility of quantifying the uncertainty of theoretical one-neutron separation energies ($S_{1n}$) when a variety of mass models are available, using the Ensemble Baysian Model Averaging (EBMA) method. 
\textcolor{black}{
The EBMA method creates a weighted average of theoretical mass models as a mixture of normal distributions, whose weights and standard deviations are estimated by MCMC using the No-U-Turn-Sampler (NUTS) from experimental $S_{1n}$ data. The resulting weights and standard deviations are expressed as distributions. The distributions of the EBMA parameters are then used to estimate the uncertainty in $S_{1n}$.
}

EBMA models have been constructed in four different ways of fitting the experimental data, namely, the whole chart of nuclides, each isotopic chain, each isotonic chain, and each isobaric chain. 
\textcolor{black}{
The mass models included in the ensemble were WS4, DZ29, KUTY05, HFB31, and ETFSI2, which are some of the most popular choices of mass models in $r$-process studies. For the whole chart EBMA model, WS4 was shown to have a dominant contribution.
}

\textcolor{black}{
Quality of uncertainty estimates has been examined by constructing EBMA models with the data from the AME2003 and testing the extrapolation performance on the new data from the AME2020. Our analysis shows that the whole chart EBMA model provides reliable uncertainty estimates. Furthermore, it was found that the $S_{1n}$ values of some of the nuclei on or close to the $N=Z$ line significantly deviate from the uncertainty estimates, suggesting that the mass models in the ensemble do not fully capture the Wigner effect. 
}

\textcolor{black}{
Although the EBMA models constructed with the data for each isotopic, isotonic, and isobaric chain were expected to better capture the local trends of the $S_{1n}$ values, they exhibit signs of underconfidence and overfitting. However, the weights in the EBMA models for individual chains, especially the isotopic and isotonic EBMA models, show which mass models perform well in different parts of the chart of nuclides, potentially giving insight into what theoretical descriptions of nuclei are effective in different regions.
}

\textcolor{black}{
Effect of linear bias correction of each mass model was examined, which was found to facilitate the overfitting of the EBMA models constructed for individual chains, and has little effect on the whole chart EBMA model. Therefore, linear bias correction is not recommended in general.
}

\textcolor{black}{
Effect of choices of mass models included in the ensemble was investigated by removing WS4, which was the dominant model in the original ensemble. It was found that the quality of uncertainty estimates provided by the whole chart EBMA model slightly degrades, but still decent. On the basis of the results and the general structure of the EBMA, it is recommended that a wide range of mass models be included for more reliable and constrained extrapolation of uncertainty estimates.
}

\textcolor{black}{
We note the performance of EBMA on extrapolation of uncertainty estimates has only been tested with the new data in the AME2020, which lie just outside the AME2003 data. In order for the EBMA method to gain more validity in the neutron-rich region, improvements of the method that put more emphasis on the neutron-rich region will be necessary. Alternatively, inferring the weights from the $r$-process observables may be considered, since the $r$-process is known to involve extremely neutron rich nuclei.
}

\begin{acknowledgments}
\textcolor{black}{
We thank S. Yoshida for a helpful discussion. \textcolor{black}{We also thank the anonymous referee for their thorough and constructive feedback, which contributed to the improvement of the manusctipt.}
Y.S. and I.D. acknowledge funding from the Canadian Natural Sciences and Engineering Research Council (NSERC), NSERC Discovery Grant No.~SAPIN-2019-00030, No.~SAPPJ-2017-00026, and the NSERC CREATE Program IsoSiM (Isotopes for Science and Medicine).
\textcolor{black}{R.K. is supported by the U.S. Department of Energy, Office of Science, Office of Nuclear Physics under Contract No. DE-AC02-05CH11231.}
M.R.M. is supported by the U.S. Department of Energy
through the Los Alamos National Laboratory (LANL). LANL
is operated by Triad National Security LLC for the National
Nuclear Security Administration of the U.S. Department of
Energy (Contract No. 89233218CNA000001).
\textcolor{black}{Y.S. and R.S. acknowledges support from the U.S. National Science Foundation under grant number 21-16686 (NP3M).
}
R.S. acknowledges support from the U.S. Department of Energy contract DE-FG02-95-ER40934.}

\end{acknowledgments}
\bibliography{main}

\begin{thebibliography}{46}%
\makeatletter
\providecommand \@ifxundefined [1]{%
 \@ifx{#1\undefined}
}%
\providecommand \@ifnum [1]{%
 \ifnum #1\expandafter \@firstoftwo
 \else \expandafter \@secondoftwo
 \fi
}%
\providecommand \@ifx [1]{%
 \ifx #1\expandafter \@firstoftwo
 \else \expandafter \@secondoftwo
 \fi
}%
\providecommand \natexlab [1]{#1}%
\providecommand \enquote  [1]{``#1''}%
\providecommand \bibnamefont  [1]{#1}%
\providecommand \bibfnamefont [1]{#1}%
\providecommand \citenamefont [1]{#1}%
\providecommand \href@noop [0]{\@secondoftwo}%
\providecommand \href [0]{\begingroup \@sanitize@url \@href}%
\providecommand \@href[1]{\@@startlink{#1}\@@href}%
\providecommand \@@href[1]{\endgroup#1\@@endlink}%
\providecommand \@sanitize@url [0]{\catcode `\\12\catcode `\$12\catcode
  `\&12\catcode `\#12\catcode `\^12\catcode `\_12\catcode `\%12\relax}%
\providecommand \@@startlink[1]{}%
\providecommand \@@endlink[0]{}%
\providecommand \url  [0]{\begingroup\@sanitize@url \@url }%
\providecommand \@url [1]{\endgroup\@href {#1}{\urlprefix }}%
\providecommand \urlprefix  [0]{URL }%
\providecommand \Eprint [0]{\href }%
\providecommand \doibase [0]{https://doi.org/}%
\providecommand \selectlanguage [0]{\@gobble}%
\providecommand \bibinfo  [0]{\@secondoftwo}%
\providecommand \bibfield  [0]{\@secondoftwo}%
\providecommand \translation [1]{[#1]}%
\providecommand \BibitemOpen [0]{}%
\providecommand \bibitemStop [0]{}%
\providecommand \bibitemNoStop [0]{.\EOS\space}%
\providecommand \EOS [0]{\spacefactor3000\relax}%
\providecommand \BibitemShut  [1]{\csname bibitem#1\endcsname}%
\let\auto@bib@innerbib\@empty
\bibitem [{\citenamefont {Moller}\ \emph {et~al.}(1995)\citenamefont {Moller},
  \citenamefont {Nix}, \citenamefont {Myers},\ and\ \citenamefont
  {Swiatecki}}]{Moller1995}%
  \BibitemOpen
  \bibfield  {author} {\bibinfo {author} {\bibfnamefont {P.}~\bibnamefont
  {Moller}}, \bibinfo {author} {\bibfnamefont {J.}~\bibnamefont {Nix}},
  \bibinfo {author} {\bibfnamefont {W.}~\bibnamefont {Myers}},\ and\ \bibinfo
  {author} {\bibfnamefont {W.}~\bibnamefont {Swiatecki}},\ }\href
  {https://doi.org/https://doi.org/10.1006/adnd.1995.1002} {\bibfield
  {journal} {\bibinfo  {journal} {Atomic Data and Nuclear Data Tables}\
  }\textbf {\bibinfo {volume} {59}},\ \bibinfo {pages} {185} (\bibinfo {year}
  {1995})}\BibitemShut {NoStop}%
\bibitem [{\citenamefont {Möller}\ \emph {et~al.}(2016)\citenamefont
  {Möller}, \citenamefont {Sierk}, \citenamefont {Ichikawa},\ and\
  \citenamefont {Sagawa}}]{Moller2016}%
  \BibitemOpen
  \bibfield  {author} {\bibinfo {author} {\bibfnamefont {P.}~\bibnamefont
  {Möller}}, \bibinfo {author} {\bibfnamefont {A.}~\bibnamefont {Sierk}},
  \bibinfo {author} {\bibfnamefont {T.}~\bibnamefont {Ichikawa}},\ and\
  \bibinfo {author} {\bibfnamefont {H.}~\bibnamefont {Sagawa}},\ }\href
  {https://doi.org/https://doi.org/10.1016/j.adt.2015.10.002} {\bibfield
  {journal} {\bibinfo  {journal} {Atomic Data and Nuclear Data Tables}\
  }\textbf {\bibinfo {volume} {109-110}},\ \bibinfo {pages} {1} (\bibinfo
  {year} {2016})}\BibitemShut {NoStop}%
\bibitem [{\citenamefont {Wang}\ \emph
  {et~al.}(2010{\natexlab{a}})\citenamefont {Wang}, \citenamefont {Liu},\ and\
  \citenamefont {Wu}}]{Wang2010a}%
  \BibitemOpen
  \bibfield  {author} {\bibinfo {author} {\bibfnamefont {N.}~\bibnamefont
  {Wang}}, \bibinfo {author} {\bibfnamefont {M.}~\bibnamefont {Liu}},\ and\
  \bibinfo {author} {\bibfnamefont {X.}~\bibnamefont {Wu}},\ }\href
  {https://doi.org/10.1103/PhysRevC.81.044322} {\bibfield  {journal} {\bibinfo
  {journal} {Phys. Rev. C}\ }\textbf {\bibinfo {volume} {81}},\ \bibinfo
  {pages} {044322} (\bibinfo {year} {2010}{\natexlab{a}})}\BibitemShut
  {NoStop}%
\bibitem [{\citenamefont {Wang}\ \emph
  {et~al.}(2010{\natexlab{b}})\citenamefont {Wang}, \citenamefont {Liang},
  \citenamefont {Liu},\ and\ \citenamefont {Wu}}]{Wang2010b}%
  \BibitemOpen
  \bibfield  {author} {\bibinfo {author} {\bibfnamefont {N.}~\bibnamefont
  {Wang}}, \bibinfo {author} {\bibfnamefont {Z.}~\bibnamefont {Liang}},
  \bibinfo {author} {\bibfnamefont {M.}~\bibnamefont {Liu}},\ and\ \bibinfo
  {author} {\bibfnamefont {X.}~\bibnamefont {Wu}},\ }\href
  {https://doi.org/10.1103/PhysRevC.82.044304} {\bibfield  {journal} {\bibinfo
  {journal} {Phys. Rev. C}\ }\textbf {\bibinfo {volume} {82}},\ \bibinfo
  {pages} {044304} (\bibinfo {year} {2010}{\natexlab{b}})}\BibitemShut
  {NoStop}%
\bibitem [{\citenamefont {Liu}\ \emph {et~al.}(2011)\citenamefont {Liu},
  \citenamefont {Wang}, \citenamefont {Deng},\ and\ \citenamefont
  {Wu}}]{Liu2011}%
  \BibitemOpen
  \bibfield  {author} {\bibinfo {author} {\bibfnamefont {M.}~\bibnamefont
  {Liu}}, \bibinfo {author} {\bibfnamefont {N.}~\bibnamefont {Wang}}, \bibinfo
  {author} {\bibfnamefont {Y.}~\bibnamefont {Deng}},\ and\ \bibinfo {author}
  {\bibfnamefont {X.}~\bibnamefont {Wu}},\ }\href
  {https://doi.org/10.1103/PhysRevC.84.014333} {\bibfield  {journal} {\bibinfo
  {journal} {Phys. Rev. C}\ }\textbf {\bibinfo {volume} {84}},\ \bibinfo
  {pages} {014333} (\bibinfo {year} {2011})}\BibitemShut {NoStop}%
\bibitem [{\citenamefont {Wang}\ \emph {et~al.}(2014)\citenamefont {Wang},
  \citenamefont {Liu}, \citenamefont {Wu},\ and\ \citenamefont
  {Meng}}]{Wang2014}%
  \BibitemOpen
  \bibfield  {author} {\bibinfo {author} {\bibfnamefont {N.}~\bibnamefont
  {Wang}}, \bibinfo {author} {\bibfnamefont {M.}~\bibnamefont {Liu}}, \bibinfo
  {author} {\bibfnamefont {X.}~\bibnamefont {Wu}},\ and\ \bibinfo {author}
  {\bibfnamefont {J.}~\bibnamefont {Meng}},\ }\href
  {https://doi.org/https://doi.org/10.1016/j.physletb.2014.05.049} {\bibfield
  {journal} {\bibinfo  {journal} {Physics Letters B}\ }\textbf {\bibinfo
  {volume} {734}},\ \bibinfo {pages} {215} (\bibinfo {year}
  {2014})}\BibitemShut {NoStop}%
\bibitem [{\citenamefont {Duflo}\ and\ \citenamefont
  {Zuker}(1995)}]{Duflo1995}%
  \BibitemOpen
  \bibfield  {author} {\bibinfo {author} {\bibfnamefont {J.}~\bibnamefont
  {Duflo}}\ and\ \bibinfo {author} {\bibfnamefont {A.~P.}\ \bibnamefont
  {Zuker}},\ }\href {https://doi.org/10.1103/PhysRevC.52.R23} {\bibfield
  {journal} {\bibinfo  {journal} {Phys. Rev. C}\ }\textbf {\bibinfo {volume}
  {52}},\ \bibinfo {pages} {R23} (\bibinfo {year} {1995})}\BibitemShut
  {NoStop}%
\bibitem [{\citenamefont {Erler}\ \emph {et~al.}(2012)\citenamefont {Erler},
  \citenamefont {Birge}, \citenamefont {Kortelainen}, \citenamefont
  {Nazarewicz}, \citenamefont {Olsen}, \citenamefont {Perhac},\ and\
  \citenamefont {Stoitsov}}]{Erler2012}%
  \BibitemOpen
  \bibfield  {author} {\bibinfo {author} {\bibfnamefont {J.}~\bibnamefont
  {Erler}}, \bibinfo {author} {\bibfnamefont {N.}~\bibnamefont {Birge}},
  \bibinfo {author} {\bibfnamefont {M.}~\bibnamefont {Kortelainen}}, \bibinfo
  {author} {\bibfnamefont {W.}~\bibnamefont {Nazarewicz}}, \bibinfo {author}
  {\bibfnamefont {E.}~\bibnamefont {Olsen}}, \bibinfo {author} {\bibfnamefont
  {A.~M.}\ \bibnamefont {Perhac}},\ and\ \bibinfo {author} {\bibfnamefont
  {M.}~\bibnamefont {Stoitsov}},\ }\href@noop {} {\bibfield  {journal}
  {\bibinfo  {journal} {Nature}\ }\textbf {\bibinfo {volume} {486}},\ \bibinfo
  {pages} {509} (\bibinfo {year} {2012})}\BibitemShut {NoStop}%
\bibitem [{\citenamefont {Goriely}\ \emph {et~al.}(2009)\citenamefont
  {Goriely}, \citenamefont {Hilaire}, \citenamefont {Girod},\ and\
  \citenamefont {P\'eru}}]{Goriely2009}%
  \BibitemOpen
  \bibfield  {author} {\bibinfo {author} {\bibfnamefont {S.}~\bibnamefont
  {Goriely}}, \bibinfo {author} {\bibfnamefont {S.}~\bibnamefont {Hilaire}},
  \bibinfo {author} {\bibfnamefont {M.}~\bibnamefont {Girod}},\ and\ \bibinfo
  {author} {\bibfnamefont {S.}~\bibnamefont {P\'eru}},\ }\href
  {https://doi.org/10.1103/PhysRevLett.102.242501} {\bibfield  {journal}
  {\bibinfo  {journal} {Phys. Rev. Lett.}\ }\textbf {\bibinfo {volume} {102}},\
  \bibinfo {pages} {242501} (\bibinfo {year} {2009})}\BibitemShut {NoStop}%
\bibitem [{\citenamefont {Goriely}\ \emph {et~al.}(2016)\citenamefont
  {Goriely}, \citenamefont {Chamel},\ and\ \citenamefont
  {Pearson}}]{Goriely2016}%
  \BibitemOpen
  \bibfield  {author} {\bibinfo {author} {\bibfnamefont {S.}~\bibnamefont
  {Goriely}}, \bibinfo {author} {\bibfnamefont {N.}~\bibnamefont {Chamel}},\
  and\ \bibinfo {author} {\bibfnamefont {J.~M.}\ \bibnamefont {Pearson}},\
  }\href {https://doi.org/10.1103/PhysRevC.93.034337} {\bibfield  {journal}
  {\bibinfo  {journal} {Phys. Rev. C}\ }\textbf {\bibinfo {volume} {93}},\
  \bibinfo {pages} {034337} (\bibinfo {year} {2016})}\BibitemShut {NoStop}%
\bibitem [{\citenamefont {Mendoza-Temis}\ \emph {et~al.}(2015)\citenamefont
  {Mendoza-Temis}, \citenamefont {Wu}, \citenamefont {Langanke}, \citenamefont
  {Mart\'{\i}nez-Pinedo}, \citenamefont {Bauswein},\ and\ \citenamefont
  {Janka}}]{MendozaTemis2015}%
  \BibitemOpen
  \bibfield  {author} {\bibinfo {author} {\bibfnamefont {J.~d.~J.}\
  \bibnamefont {Mendoza-Temis}}, \bibinfo {author} {\bibfnamefont {M.-R.}\
  \bibnamefont {Wu}}, \bibinfo {author} {\bibfnamefont {K.}~\bibnamefont
  {Langanke}}, \bibinfo {author} {\bibfnamefont {G.}~\bibnamefont
  {Mart\'{\i}nez-Pinedo}}, \bibinfo {author} {\bibfnamefont {A.}~\bibnamefont
  {Bauswein}},\ and\ \bibinfo {author} {\bibfnamefont {H.-T.}\ \bibnamefont
  {Janka}},\ }\href {https://doi.org/10.1103/PhysRevC.92.055805} {\bibfield
  {journal} {\bibinfo  {journal} {Phys. Rev. C}\ }\textbf {\bibinfo {volume}
  {92}},\ \bibinfo {pages} {055805} (\bibinfo {year} {2015})}\BibitemShut
  {NoStop}%
\bibitem [{\citenamefont {Martin}\ \emph {et~al.}(2016)\citenamefont {Martin},
  \citenamefont {Arcones}, \citenamefont {Nazarewicz},\ and\ \citenamefont
  {Olsen}}]{Martin2016}%
  \BibitemOpen
  \bibfield  {author} {\bibinfo {author} {\bibfnamefont {D.}~\bibnamefont
  {Martin}}, \bibinfo {author} {\bibfnamefont {A.}~\bibnamefont {Arcones}},
  \bibinfo {author} {\bibfnamefont {W.}~\bibnamefont {Nazarewicz}},\ and\
  \bibinfo {author} {\bibfnamefont {E.}~\bibnamefont {Olsen}},\ }\href
  {https://doi.org/10.1103/PhysRevLett.116.121101} {\bibfield  {journal}
  {\bibinfo  {journal} {Phys. Rev. Lett.}\ }\textbf {\bibinfo {volume} {116}},\
  \bibinfo {pages} {121101} (\bibinfo {year} {2016})}\BibitemShut {NoStop}%
\bibitem [{\citenamefont {Mumpower}\ \emph {et~al.}(2016)\citenamefont
  {Mumpower}, \citenamefont {Surman}, \citenamefont {McLaughlin},\ and\
  \citenamefont {Aprahamian}}]{Mumpower2016}%
  \BibitemOpen
  \bibfield  {author} {\bibinfo {author} {\bibfnamefont {M.}~\bibnamefont
  {Mumpower}}, \bibinfo {author} {\bibfnamefont {R.}~\bibnamefont {Surman}},
  \bibinfo {author} {\bibfnamefont {G.}~\bibnamefont {McLaughlin}},\ and\
  \bibinfo {author} {\bibfnamefont {A.}~\bibnamefont {Aprahamian}},\ }\href
  {https://doi.org/https://doi.org/10.1016/j.ppnp.2015.09.001} {\bibfield
  {journal} {\bibinfo  {journal} {Progress in Particle and Nuclear Physics}\
  }\textbf {\bibinfo {volume} {86}},\ \bibinfo {pages} {86} (\bibinfo {year}
  {2016})}\BibitemShut {NoStop}%
\bibitem [{\citenamefont {Zhu}\ \emph {et~al.}(2021)\citenamefont {Zhu},
  \citenamefont {Lund}, \citenamefont {Barnes}, \citenamefont {Sprouse},
  \citenamefont {Vassh}, \citenamefont {McLaughlin}, \citenamefont {Mumpower},\
  and\ \citenamefont {Surman}}]{Zhu2021}%
  \BibitemOpen
  \bibfield  {author} {\bibinfo {author} {\bibfnamefont {Y.~L.}\ \bibnamefont
  {Zhu}}, \bibinfo {author} {\bibfnamefont {K.~A.}\ \bibnamefont {Lund}},
  \bibinfo {author} {\bibfnamefont {J.}~\bibnamefont {Barnes}}, \bibinfo
  {author} {\bibfnamefont {T.~M.}\ \bibnamefont {Sprouse}}, \bibinfo {author}
  {\bibfnamefont {N.}~\bibnamefont {Vassh}}, \bibinfo {author} {\bibfnamefont
  {G.~C.}\ \bibnamefont {McLaughlin}}, \bibinfo {author} {\bibfnamefont
  {M.~R.}\ \bibnamefont {Mumpower}},\ and\ \bibinfo {author} {\bibfnamefont
  {R.}~\bibnamefont {Surman}},\ }\href
  {https://doi.org/10.3847/1538-4357/abc69e} {\bibfield  {journal} {\bibinfo
  {journal} {The Astrophysical Journal}\ }\textbf {\bibinfo {volume} {906}},\
  \bibinfo {pages} {94} (\bibinfo {year} {2021})}\BibitemShut {NoStop}%
\bibitem [{\citenamefont {Barnes}\ \emph {et~al.}(2021)\citenamefont {Barnes},
  \citenamefont {Zhu}, \citenamefont {Lund}, \citenamefont {Sprouse},
  \citenamefont {Vassh}, \citenamefont {McLaughlin}, \citenamefont {Mumpower},\
  and\ \citenamefont {Surman}}]{Barnes2021}%
  \BibitemOpen
  \bibfield  {author} {\bibinfo {author} {\bibfnamefont {J.}~\bibnamefont
  {Barnes}}, \bibinfo {author} {\bibfnamefont {Y.~L.}\ \bibnamefont {Zhu}},
  \bibinfo {author} {\bibfnamefont {K.~A.}\ \bibnamefont {Lund}}, \bibinfo
  {author} {\bibfnamefont {T.~M.}\ \bibnamefont {Sprouse}}, \bibinfo {author}
  {\bibfnamefont {N.}~\bibnamefont {Vassh}}, \bibinfo {author} {\bibfnamefont
  {G.~C.}\ \bibnamefont {McLaughlin}}, \bibinfo {author} {\bibfnamefont
  {M.~R.}\ \bibnamefont {Mumpower}},\ and\ \bibinfo {author} {\bibfnamefont
  {R.}~\bibnamefont {Surman}},\ }\href
  {https://doi.org/10.3847/1538-4357/ac0aec} {\bibfield  {journal} {\bibinfo
  {journal} {The Astrophysical Journal}\ }\textbf {\bibinfo {volume} {918}},\
  \bibinfo {pages} {44} (\bibinfo {year} {2021})}\BibitemShut {NoStop}%
\bibitem [{\citenamefont {Dobaczewski}\ \emph {et~al.}(2014)\citenamefont
  {Dobaczewski}, \citenamefont {Nazarewicz},\ and\ \citenamefont
  {Reinhard}}]{Dobaczewski2014}%
  \BibitemOpen
  \bibfield  {author} {\bibinfo {author} {\bibfnamefont {J.}~\bibnamefont
  {Dobaczewski}}, \bibinfo {author} {\bibfnamefont {W.}~\bibnamefont
  {Nazarewicz}},\ and\ \bibinfo {author} {\bibfnamefont {P.-G.}\ \bibnamefont
  {Reinhard}},\ }\href {https://doi.org/10.1088/0954-3899/41/7/074001}
  {\bibfield  {journal} {\bibinfo  {journal} {Journal of Physics G: Nuclear and
  Particle Physics}\ }\textbf {\bibinfo {volume} {41}},\ \bibinfo {pages}
  {074001} (\bibinfo {year} {2014})}\BibitemShut {NoStop}%
\bibitem [{\citenamefont {McDonnell}\ \emph {et~al.}(2015)\citenamefont
  {McDonnell}, \citenamefont {Schunck}, \citenamefont {Higdon}, \citenamefont
  {Sarich}, \citenamefont {Wild},\ and\ \citenamefont
  {Nazarewicz}}]{McDonnell2015}%
  \BibitemOpen
  \bibfield  {author} {\bibinfo {author} {\bibfnamefont {J.~D.}\ \bibnamefont
  {McDonnell}}, \bibinfo {author} {\bibfnamefont {N.}~\bibnamefont {Schunck}},
  \bibinfo {author} {\bibfnamefont {D.}~\bibnamefont {Higdon}}, \bibinfo
  {author} {\bibfnamefont {J.}~\bibnamefont {Sarich}}, \bibinfo {author}
  {\bibfnamefont {S.~M.}\ \bibnamefont {Wild}},\ and\ \bibinfo {author}
  {\bibfnamefont {W.}~\bibnamefont {Nazarewicz}},\ }\href
  {https://doi.org/10.1103/PhysRevLett.114.122501} {\bibfield  {journal}
  {\bibinfo  {journal} {Phys. Rev. Lett.}\ }\textbf {\bibinfo {volume} {114}},\
  \bibinfo {pages} {122501} (\bibinfo {year} {2015})}\BibitemShut {NoStop}%
\bibitem [{\citenamefont {Wang}\ \emph {et~al.}(2021)\citenamefont {Wang},
  \citenamefont {Huang}, \citenamefont {Kondev}, \citenamefont {Audi},\ and\
  \citenamefont {Naimi}}]{Wang2021}%
  \BibitemOpen
  \bibfield  {author} {\bibinfo {author} {\bibfnamefont {M.}~\bibnamefont
  {Wang}}, \bibinfo {author} {\bibfnamefont {W.}~\bibnamefont {Huang}},
  \bibinfo {author} {\bibfnamefont {F.}~\bibnamefont {Kondev}}, \bibinfo
  {author} {\bibfnamefont {G.}~\bibnamefont {Audi}},\ and\ \bibinfo {author}
  {\bibfnamefont {S.}~\bibnamefont {Naimi}},\ }\href
  {https://doi.org/10.1088/1674-1137/abddaf} {\bibfield  {journal} {\bibinfo
  {journal} {Chinese Physics C}\ }\textbf {\bibinfo {volume} {45}},\ \bibinfo
  {pages} {030003} (\bibinfo {year} {2021})}\BibitemShut {NoStop}%
\bibitem [{\citenamefont {Raftery}\ \emph {et~al.}(2005)\citenamefont
  {Raftery}, \citenamefont {Gneiting}, \citenamefont {Balabdaoui},\ and\
  \citenamefont {Polakowski}}]{Raftery2005}%
  \BibitemOpen
  \bibfield  {author} {\bibinfo {author} {\bibfnamefont {A.~E.}\ \bibnamefont
  {Raftery}}, \bibinfo {author} {\bibfnamefont {T.}~\bibnamefont {Gneiting}},
  \bibinfo {author} {\bibfnamefont {F.}~\bibnamefont {Balabdaoui}},\ and\
  \bibinfo {author} {\bibfnamefont {M.}~\bibnamefont {Polakowski}},\ }\href
  {https://doi.org/10.1175/MWR2906.1} {\bibfield  {journal} {\bibinfo
  {journal} {Monthly Weather Review}\ }\textbf {\bibinfo {volume} {133}},\
  \bibinfo {pages} {1155 } (\bibinfo {year} {2005})}\BibitemShut {NoStop}%
\bibitem [{\citenamefont {Niu}\ and\ \citenamefont {Liang}(2022)}]{Niu2022}%
  \BibitemOpen
  \bibfield  {author} {\bibinfo {author} {\bibfnamefont {Z.~M.}\ \bibnamefont
  {Niu}}\ and\ \bibinfo {author} {\bibfnamefont {H.~Z.}\ \bibnamefont
  {Liang}},\ }\href {https://doi.org/10.1103/PhysRevC.106.L021303} {\bibfield
  {journal} {\bibinfo  {journal} {Phys. Rev. C}\ }\textbf {\bibinfo {volume}
  {106}},\ \bibinfo {pages} {L021303} (\bibinfo {year} {2022})}\BibitemShut
  {NoStop}%
\bibitem [{\citenamefont {Niu}\ and\ \citenamefont {Liang}(2018)}]{Niu2018}%
  \BibitemOpen
  \bibfield  {author} {\bibinfo {author} {\bibfnamefont {Z.}~\bibnamefont
  {Niu}}\ and\ \bibinfo {author} {\bibfnamefont {H.}~\bibnamefont {Liang}},\
  }\href {https://doi.org/https://doi.org/10.1016/j.physletb.2018.01.002}
  {\bibfield  {journal} {\bibinfo  {journal} {Physics Letters B}\ }\textbf
  {\bibinfo {volume} {778}},\ \bibinfo {pages} {48} (\bibinfo {year}
  {2018})}\BibitemShut {NoStop}%
\bibitem [{\citenamefont {Neufcourt}\ \emph {et~al.}(2018)\citenamefont
  {Neufcourt}, \citenamefont {Cao}, \citenamefont {Nazarewicz},\ and\
  \citenamefont {Viens}}]{Neufcourt2018}%
  \BibitemOpen
  \bibfield  {author} {\bibinfo {author} {\bibfnamefont {L.}~\bibnamefont
  {Neufcourt}}, \bibinfo {author} {\bibfnamefont {Y.}~\bibnamefont {Cao}},
  \bibinfo {author} {\bibfnamefont {W.}~\bibnamefont {Nazarewicz}},\ and\
  \bibinfo {author} {\bibfnamefont {F.}~\bibnamefont {Viens}},\ }\href
  {https://doi.org/10.1103/PhysRevC.98.034318} {\bibfield  {journal} {\bibinfo
  {journal} {Phys. Rev. C}\ }\textbf {\bibinfo {volume} {98}},\ \bibinfo
  {pages} {034318} (\bibinfo {year} {2018})}\BibitemShut {NoStop}%
\bibitem [{\citenamefont {Neufcourt}\ \emph {et~al.}(2019)\citenamefont
  {Neufcourt}, \citenamefont {Cao}, \citenamefont {Nazarewicz}, \citenamefont
  {Olsen},\ and\ \citenamefont {Viens}}]{Neufcourt2019}%
  \BibitemOpen
  \bibfield  {author} {\bibinfo {author} {\bibfnamefont {L.}~\bibnamefont
  {Neufcourt}}, \bibinfo {author} {\bibfnamefont {Y.}~\bibnamefont {Cao}},
  \bibinfo {author} {\bibfnamefont {W.}~\bibnamefont {Nazarewicz}}, \bibinfo
  {author} {\bibfnamefont {E.}~\bibnamefont {Olsen}},\ and\ \bibinfo {author}
  {\bibfnamefont {F.}~\bibnamefont {Viens}},\ }\href
  {https://doi.org/10.1103/PhysRevLett.122.062502} {\bibfield  {journal}
  {\bibinfo  {journal} {Phys. Rev. Lett.}\ }\textbf {\bibinfo {volume} {122}},\
  \bibinfo {pages} {062502} (\bibinfo {year} {2019})}\BibitemShut {NoStop}%
\bibitem [{\citenamefont {Lasseri}\ \emph {et~al.}(2020)\citenamefont
  {Lasseri}, \citenamefont {Regnier}, \citenamefont {Ebran},\ and\
  \citenamefont {Penon}}]{Lasseri2020}%
  \BibitemOpen
  \bibfield  {author} {\bibinfo {author} {\bibfnamefont {R.-D.}\ \bibnamefont
  {Lasseri}}, \bibinfo {author} {\bibfnamefont {D.}~\bibnamefont {Regnier}},
  \bibinfo {author} {\bibfnamefont {J.-P.}\ \bibnamefont {Ebran}},\ and\
  \bibinfo {author} {\bibfnamefont {A.}~\bibnamefont {Penon}},\ }\href
  {https://doi.org/10.1103/PhysRevLett.124.162502} {\bibfield  {journal}
  {\bibinfo  {journal} {Phys. Rev. Lett.}\ }\textbf {\bibinfo {volume} {124}},\
  \bibinfo {pages} {162502} (\bibinfo {year} {2020})}\BibitemShut {NoStop}%
\bibitem [{\citenamefont {Utama}\ \emph {et~al.}(2016)\citenamefont {Utama},
  \citenamefont {Piekarewicz},\ and\ \citenamefont {Prosper}}]{Utama2016}%
  \BibitemOpen
  \bibfield  {author} {\bibinfo {author} {\bibfnamefont {R.}~\bibnamefont
  {Utama}}, \bibinfo {author} {\bibfnamefont {J.}~\bibnamefont {Piekarewicz}},\
  and\ \bibinfo {author} {\bibfnamefont {H.~B.}\ \bibnamefont {Prosper}},\
  }\href {https://doi.org/10.1103/PhysRevC.93.014311} {\bibfield  {journal}
  {\bibinfo  {journal} {Phys. Rev. C}\ }\textbf {\bibinfo {volume} {93}},\
  \bibinfo {pages} {014311} (\bibinfo {year} {2016})}\BibitemShut {NoStop}%
\bibitem [{\citenamefont {Lovell}\ \emph {et~al.}(2022)\citenamefont {Lovell},
  \citenamefont {Mohan}, \citenamefont {Sprouse},\ and\ \citenamefont
  {Mumpower}}]{Lovell2022}%
  \BibitemOpen
  \bibfield  {author} {\bibinfo {author} {\bibfnamefont {A.~E.}\ \bibnamefont
  {Lovell}}, \bibinfo {author} {\bibfnamefont {A.~T.}\ \bibnamefont {Mohan}},
  \bibinfo {author} {\bibfnamefont {T.~M.}\ \bibnamefont {Sprouse}},\ and\
  \bibinfo {author} {\bibfnamefont {M.~R.}\ \bibnamefont {Mumpower}},\ }\href
  {https://doi.org/10.1103/PhysRevC.106.014305} {\bibfield  {journal} {\bibinfo
   {journal} {Phys. Rev. C}\ }\textbf {\bibinfo {volume} {106}},\ \bibinfo
  {pages} {014305} (\bibinfo {year} {2022})}\BibitemShut {NoStop}%
\bibitem [{\citenamefont {Mumpower}\ \emph {et~al.}(2022)\citenamefont
  {Mumpower}, \citenamefont {Sprouse}, \citenamefont {Lovell},\ and\
  \citenamefont {Mohan}}]{Mumpower2022}%
  \BibitemOpen
  \bibfield  {author} {\bibinfo {author} {\bibfnamefont {M.~R.}\ \bibnamefont
  {Mumpower}}, \bibinfo {author} {\bibfnamefont {T.~M.}\ \bibnamefont
  {Sprouse}}, \bibinfo {author} {\bibfnamefont {A.~E.}\ \bibnamefont
  {Lovell}},\ and\ \bibinfo {author} {\bibfnamefont {A.~T.}\ \bibnamefont
  {Mohan}},\ }\href {https://doi.org/10.1103/PhysRevC.106.L021301} {\bibfield
  {journal} {\bibinfo  {journal} {Phys. Rev. C}\ }\textbf {\bibinfo {volume}
  {106}},\ \bibinfo {pages} {L021301} (\bibinfo {year} {2022})}\BibitemShut
  {NoStop}%
\bibitem [{\citenamefont {Mumpower}\ \emph {et~al.}(2023)\citenamefont
  {Mumpower}, \citenamefont {Li}, \citenamefont {Sprouse}, \citenamefont
  {Meyer}, \citenamefont {Lovell},\ and\ \citenamefont {Mohan}}]{Mumpower2023}%
  \BibitemOpen
  \bibfield  {author} {\bibinfo {author} {\bibfnamefont {M.}~\bibnamefont
  {Mumpower}}, \bibinfo {author} {\bibfnamefont {M.}~\bibnamefont {Li}},
  \bibinfo {author} {\bibfnamefont {T.~M.}\ \bibnamefont {Sprouse}}, \bibinfo
  {author} {\bibfnamefont {B.~S.}\ \bibnamefont {Meyer}}, \bibinfo {author}
  {\bibfnamefont {A.~E.}\ \bibnamefont {Lovell}},\ and\ \bibinfo {author}
  {\bibfnamefont {A.~T.}\ \bibnamefont {Mohan}},\ }\bibfield  {journal}
  {\bibinfo  {journal} {Frontiers in Physics}\ }\textbf {\bibinfo {volume}
  {11}},\ \href {https://doi.org/10.3389/fphy.2023.1198572}
  {10.3389/fphy.2023.1198572} (\bibinfo {year} {2023})\BibitemShut {NoStop}%
\bibitem [{\citenamefont {Hoeting}\ \emph {et~al.}(1999)\citenamefont
  {Hoeting}, \citenamefont {Madigan}, \citenamefont {Raftery},\ and\
  \citenamefont {Volinsky}}]{Hoeting1999}%
  \BibitemOpen
  \bibfield  {author} {\bibinfo {author} {\bibfnamefont {J.~A.}\ \bibnamefont
  {Hoeting}}, \bibinfo {author} {\bibfnamefont {D.}~\bibnamefont {Madigan}},
  \bibinfo {author} {\bibfnamefont {A.~E.}\ \bibnamefont {Raftery}},\ and\
  \bibinfo {author} {\bibfnamefont {C.~T.}\ \bibnamefont {Volinsky}},\ }\href
  {https://doi.org/10.1214/ss/1009212519} {\bibfield  {journal} {\bibinfo
  {journal} {Statistical Science}\ }\textbf {\bibinfo {volume} {14}},\ \bibinfo
  {pages} {382 } (\bibinfo {year} {1999})}\BibitemShut {NoStop}%
\bibitem [{\citenamefont {Gelman}\ \emph {et~al.}(2017)\citenamefont {Gelman},
  \citenamefont {Simpson},\ and\ \citenamefont
  {Betancourt}}]{Gelman_prior_2017}%
  \BibitemOpen
  \bibfield  {author} {\bibinfo {author} {\bibfnamefont {A.}~\bibnamefont
  {Gelman}}, \bibinfo {author} {\bibfnamefont {D.}~\bibnamefont {Simpson}},\
  and\ \bibinfo {author} {\bibfnamefont {M.}~\bibnamefont {Betancourt}},\
  }\href {https://www.mdpi.com/1099-4300/19/10/555} {\bibfield  {journal}
  {\bibinfo  {journal} {Entropy}\ }\textbf {\bibinfo {volume} {19}} (\bibinfo
  {year} {2017})}\BibitemShut {NoStop}%
\bibitem [{\citenamefont {Hamaker}\ \emph {et~al.}(2021)\citenamefont
  {Hamaker}, \citenamefont {Leistenschneider}, \citenamefont {Jain},
  \citenamefont {Bollen}, \citenamefont {Giuliani}, \citenamefont {Lund},
  \citenamefont {Nazarewicz}, \citenamefont {Neufcourt}, \citenamefont
  {Nicoloff}, \citenamefont {Puentes}, \citenamefont {Ringle}, \citenamefont
  {Sumithrarachchi},\ and\ \citenamefont {Yandow}}]{Hamaker2021}%
  \BibitemOpen
  \bibfield  {author} {\bibinfo {author} {\bibfnamefont {A.}~\bibnamefont
  {Hamaker}}, \bibinfo {author} {\bibfnamefont {E.}~\bibnamefont
  {Leistenschneider}}, \bibinfo {author} {\bibfnamefont {R.}~\bibnamefont
  {Jain}}, \bibinfo {author} {\bibfnamefont {G.}~\bibnamefont {Bollen}},
  \bibinfo {author} {\bibfnamefont {S.~A.}\ \bibnamefont {Giuliani}}, \bibinfo
  {author} {\bibfnamefont {K.}~\bibnamefont {Lund}}, \bibinfo {author}
  {\bibfnamefont {W.}~\bibnamefont {Nazarewicz}}, \bibinfo {author}
  {\bibfnamefont {L.}~\bibnamefont {Neufcourt}}, \bibinfo {author}
  {\bibfnamefont {C.~R.}\ \bibnamefont {Nicoloff}}, \bibinfo {author}
  {\bibfnamefont {D.}~\bibnamefont {Puentes}}, \bibinfo {author} {\bibfnamefont
  {R.}~\bibnamefont {Ringle}}, \bibinfo {author} {\bibfnamefont {C.~S.}\
  \bibnamefont {Sumithrarachchi}},\ and\ \bibinfo {author} {\bibfnamefont
  {I.~T.}\ \bibnamefont {Yandow}},\ }\href
  {https://doi.org/10.1038/s41567-021-01395-w} {\bibfield  {journal} {\bibinfo
  {journal} {Nature Physics}\ }\textbf {\bibinfo {volume} {17}},\ \bibinfo
  {pages} {1408} (\bibinfo {year} {2021})}\BibitemShut {NoStop}%
\bibitem [{\citenamefont {Neufcourt}\ \emph
  {et~al.}(2020{\natexlab{a}})\citenamefont {Neufcourt}, \citenamefont {Cao},
  \citenamefont {Giuliani}, \citenamefont {Nazarewicz}, \citenamefont {Olsen},\
  and\ \citenamefont {Tarasov}}]{Neufcourt2020}%
  \BibitemOpen
  \bibfield  {author} {\bibinfo {author} {\bibfnamefont {L.}~\bibnamefont
  {Neufcourt}}, \bibinfo {author} {\bibfnamefont {Y.}~\bibnamefont {Cao}},
  \bibinfo {author} {\bibfnamefont {S.~A.}\ \bibnamefont {Giuliani}}, \bibinfo
  {author} {\bibfnamefont {W.}~\bibnamefont {Nazarewicz}}, \bibinfo {author}
  {\bibfnamefont {E.}~\bibnamefont {Olsen}},\ and\ \bibinfo {author}
  {\bibfnamefont {O.~B.}\ \bibnamefont {Tarasov}},\ }\href
  {https://doi.org/10.1103/PhysRevC.101.044307} {\bibfield  {journal} {\bibinfo
   {journal} {Phys. Rev. C}\ }\textbf {\bibinfo {volume} {101}},\ \bibinfo
  {pages} {044307} (\bibinfo {year} {2020}{\natexlab{a}})}\BibitemShut
  {NoStop}%
\bibitem [{\citenamefont {Neufcourt}\ \emph
  {et~al.}(2020{\natexlab{b}})\citenamefont {Neufcourt}, \citenamefont {Cao},
  \citenamefont {Giuliani}, \citenamefont {Nazarewicz}, \citenamefont {Olsen},\
  and\ \citenamefont {Tarasov}}]{Neufcourt2020b}%
  \BibitemOpen
  \bibfield  {author} {\bibinfo {author} {\bibfnamefont {L.}~\bibnamefont
  {Neufcourt}}, \bibinfo {author} {\bibfnamefont {Y.}~\bibnamefont {Cao}},
  \bibinfo {author} {\bibfnamefont {S.}~\bibnamefont {Giuliani}}, \bibinfo
  {author} {\bibfnamefont {W.}~\bibnamefont {Nazarewicz}}, \bibinfo {author}
  {\bibfnamefont {E.}~\bibnamefont {Olsen}},\ and\ \bibinfo {author}
  {\bibfnamefont {O.~B.}\ \bibnamefont {Tarasov}},\ }\href
  {https://doi.org/10.1103/PhysRevC.101.014319} {\bibfield  {journal} {\bibinfo
   {journal} {Phys. Rev. C}\ }\textbf {\bibinfo {volume} {101}},\ \bibinfo
  {pages} {014319} (\bibinfo {year} {2020}{\natexlab{b}})}\BibitemShut
  {NoStop}%
\bibitem [{\citenamefont {Rasmussen}\ and\ \citenamefont
  {Williams}(2005)}]{Rasmussen2005}%
  \BibitemOpen
  \bibfield  {author} {\bibinfo {author} {\bibfnamefont {C.~E.}\ \bibnamefont
  {Rasmussen}}\ and\ \bibinfo {author} {\bibfnamefont {C.~K.~I.}\ \bibnamefont
  {Williams}},\ }\href {https://doi.org/10.7551/mitpress/3206.001.0001} {\emph
  {\bibinfo {title} {{Gaussian Processes for Machine Learning}}}}\ (\bibinfo
  {publisher} {The MIT Press},\ \bibinfo {year} {2005})\BibitemShut {NoStop}%
\bibitem [{\citenamefont {Yoshida}(2020)}]{Yoshida2020}%
  \BibitemOpen
  \bibfield  {author} {\bibinfo {author} {\bibfnamefont {S.}~\bibnamefont
  {Yoshida}},\ }\href {https://doi.org/10.1103/PhysRevC.102.024305} {\bibfield
  {journal} {\bibinfo  {journal} {Phys. Rev. C}\ }\textbf {\bibinfo {volume}
  {102}},\ \bibinfo {pages} {024305} (\bibinfo {year} {2020})}\BibitemShut
  {NoStop}%
\bibitem [{\citenamefont {Phillips}\ \emph {et~al.}(2021)\citenamefont
  {Phillips}, \citenamefont {Furnstahl}, \citenamefont {Heinz}, \citenamefont
  {Maiti}, \citenamefont {Nazarewicz}, \citenamefont {Nunes}, \citenamefont
  {Plumlee}, \citenamefont {Pratola}, \citenamefont {Pratt}, \citenamefont
  {Viens},\ and\ \citenamefont {Wild}}]{Phillips2021BAND}%
  \BibitemOpen
  \bibfield  {author} {\bibinfo {author} {\bibfnamefont {D.~R.}\ \bibnamefont
  {Phillips}}, \bibinfo {author} {\bibfnamefont {R.~J.}\ \bibnamefont
  {Furnstahl}}, \bibinfo {author} {\bibfnamefont {U.}~\bibnamefont {Heinz}},
  \bibinfo {author} {\bibfnamefont {T.}~\bibnamefont {Maiti}}, \bibinfo
  {author} {\bibfnamefont {W.}~\bibnamefont {Nazarewicz}}, \bibinfo {author}
  {\bibfnamefont {F.~M.}\ \bibnamefont {Nunes}}, \bibinfo {author}
  {\bibfnamefont {M.}~\bibnamefont {Plumlee}}, \bibinfo {author} {\bibfnamefont
  {M.~T.}\ \bibnamefont {Pratola}}, \bibinfo {author} {\bibfnamefont
  {S.}~\bibnamefont {Pratt}}, \bibinfo {author} {\bibfnamefont {F.~G.}\
  \bibnamefont {Viens}},\ and\ \bibinfo {author} {\bibfnamefont {S.~M.}\
  \bibnamefont {Wild}},\ }\href {https://doi.org/10.1088/1361-6471/abf1df}
  {\bibfield  {journal} {\bibinfo  {journal} {Journal of Physics G: Nuclear and
  Particle Physics}\ }\textbf {\bibinfo {volume} {48}},\ \bibinfo {pages}
  {072001} (\bibinfo {year} {2021})}\BibitemShut {NoStop}%
\bibitem [{\citenamefont {Semposki}\ \emph {et~al.}(2022)\citenamefont
  {Semposki}, \citenamefont {Furnstahl},\ and\ \citenamefont
  {Phillips}}]{Semposki2022BMM}%
  \BibitemOpen
  \bibfield  {author} {\bibinfo {author} {\bibfnamefont {A.~C.}\ \bibnamefont
  {Semposki}}, \bibinfo {author} {\bibfnamefont {R.~J.}\ \bibnamefont
  {Furnstahl}},\ and\ \bibinfo {author} {\bibfnamefont {D.~R.}\ \bibnamefont
  {Phillips}},\ }\href {https://doi.org/10.1103/PhysRevC.106.044002} {\bibfield
   {journal} {\bibinfo  {journal} {Phys. Rev. C}\ }\textbf {\bibinfo {volume}
  {106}},\ \bibinfo {pages} {044002} (\bibinfo {year} {2022})}\BibitemShut
  {NoStop}%
\bibitem [{\citenamefont {Salvatier}\ \emph {et~al.}(2016)\citenamefont
  {Salvatier}, \citenamefont {Wiecki},\ and\ \citenamefont
  {Fonnesbeck}}]{Salvatier2016}%
  \BibitemOpen
  \bibfield  {author} {\bibinfo {author} {\bibfnamefont {J.}~\bibnamefont
  {Salvatier}}, \bibinfo {author} {\bibfnamefont {T.~V.}\ \bibnamefont
  {Wiecki}},\ and\ \bibinfo {author} {\bibfnamefont {C.}~\bibnamefont
  {Fonnesbeck}},\ }\href {https://doi.org/10.7717/peerj-cs.55} {\bibfield
  {journal} {\bibinfo  {journal} {{PeerJ} Computer Science}\ }\textbf {\bibinfo
  {volume} {2}},\ \bibinfo {pages} {e55} (\bibinfo {year} {2016})}\BibitemShut
  {NoStop}%
\bibitem [{\citenamefont {Neal}(2012)}]{Neal_BNN_2012}%
  \BibitemOpen
  \bibfield  {author} {\bibinfo {author} {\bibfnamefont {R.~M.}\ \bibnamefont
  {Neal}},\ }\href@noop {} {\emph {\bibinfo {title} {Bayesian learning for
  neural networks}}},\ Vol.\ \bibinfo {volume} {118}\ (\bibinfo  {publisher}
  {Springer Science \& Business Media},\ \bibinfo {year} {2012})\BibitemShut
  {NoStop}%
\bibitem [{\citenamefont {Neal}\ \emph {et~al.}(2011)\citenamefont {Neal} \emph
  {et~al.}}]{Neal2011}%
  \BibitemOpen
  \bibfield  {author} {\bibinfo {author} {\bibfnamefont {R.~M.}\ \bibnamefont
  {Neal}} \emph {et~al.},\ }\href@noop {} {\bibfield  {journal} {\bibinfo
  {journal} {Handbook of markov chain monte carlo}\ }\textbf {\bibinfo {volume}
  {2}},\ \bibinfo {pages} {2} (\bibinfo {year} {2011})}\BibitemShut {NoStop}%
\bibitem [{\citenamefont {Stephens}(2000)}]{Stephens2000}%
  \BibitemOpen
  \bibfield  {author} {\bibinfo {author} {\bibfnamefont {M.}~\bibnamefont
  {Stephens}},\ }\href
  {https://doi.org/https://doi.org/10.1111/1467-9868.00265} {\bibfield
  {journal} {\bibinfo  {journal} {Journal of the Royal Statistical Society:
  Series B (Statistical Methodology)}\ }\textbf {\bibinfo {volume} {62}},\
  \bibinfo {pages} {795} (\bibinfo {year} {2000})}\BibitemShut {NoStop}%
\bibitem [{\citenamefont {Jasra}\ \emph {et~al.}(2005)\citenamefont {Jasra},
  \citenamefont {Holmes},\ and\ \citenamefont {Stephens}}]{Jasra2005}%
  \BibitemOpen
  \bibfield  {author} {\bibinfo {author} {\bibfnamefont {A.}~\bibnamefont
  {Jasra}}, \bibinfo {author} {\bibfnamefont {C.~C.}\ \bibnamefont {Holmes}},\
  and\ \bibinfo {author} {\bibfnamefont {D.~A.}\ \bibnamefont {Stephens}},\
  }\href {https://doi.org/10.1214/088342305000000016} {\bibfield  {journal}
  {\bibinfo  {journal} {Statistical Science}\ }\textbf {\bibinfo {volume}
  {20}},\ \bibinfo {pages} {50 } (\bibinfo {year} {2005})}\BibitemShut
  {NoStop}%
\bibitem [{\citenamefont {Koura}\ \emph {et~al.}(2005)\citenamefont {Koura},
  \citenamefont {Tachibana}, \citenamefont {Uno},\ and\ \citenamefont
  {Yamada}}]{Koura2005}%
  \BibitemOpen
  \bibfield  {author} {\bibinfo {author} {\bibfnamefont {H.}~\bibnamefont
  {Koura}}, \bibinfo {author} {\bibfnamefont {T.}~\bibnamefont {Tachibana}},
  \bibinfo {author} {\bibfnamefont {M.}~\bibnamefont {Uno}},\ and\ \bibinfo
  {author} {\bibfnamefont {M.}~\bibnamefont {Yamada}},\ }\href
  {https://doi.org/10.1143/PTP.113.305} {\bibfield  {journal} {\bibinfo
  {journal} {Progress of Theoretical Physics}\ }\textbf {\bibinfo {volume}
  {113}},\ \bibinfo {pages} {305} (\bibinfo {year} {2005})},\ \Eprint
  {https://arxiv.org/abs/https://academic.oup.com/ptp/article-pdf/113/2/305/5192381/113-2-305.pdf}
  {https://academic.oup.com/ptp/article-pdf/113/2/305/5192381/113-2-305.pdf}
  \BibitemShut {NoStop}%
\bibitem [{\citenamefont {Aboussir}\ \emph {et~al.}(1995)\citenamefont
  {Aboussir}, \citenamefont {Pearson}, \citenamefont {Dutta},\ and\
  \citenamefont {Tondeur}}]{Aboussir1995}%
  \BibitemOpen
  \bibfield  {author} {\bibinfo {author} {\bibfnamefont {Y.}~\bibnamefont
  {Aboussir}}, \bibinfo {author} {\bibfnamefont {J.}~\bibnamefont {Pearson}},
  \bibinfo {author} {\bibfnamefont {A.}~\bibnamefont {Dutta}},\ and\ \bibinfo
  {author} {\bibfnamefont {F.}~\bibnamefont {Tondeur}},\ }\href
  {https://doi.org/https://doi.org/10.1016/S0092-640X(95)90014-4} {\bibfield
  {journal} {\bibinfo  {journal} {Atomic Data and Nuclear Data Tables}\
  }\textbf {\bibinfo {volume} {61}},\ \bibinfo {pages} {127} (\bibinfo {year}
  {1995})}\BibitemShut {NoStop}%
\bibitem [{\citenamefont {Audi}\ \emph {et~al.}(2003)\citenamefont {Audi},
  \citenamefont {Wapstra},\ and\ \citenamefont {Thibault}}]{Audi2003}%
  \BibitemOpen
  \bibfield  {author} {\bibinfo {author} {\bibfnamefont {G.}~\bibnamefont
  {Audi}}, \bibinfo {author} {\bibfnamefont {A.}~\bibnamefont {Wapstra}},\ and\
  \bibinfo {author} {\bibfnamefont {C.}~\bibnamefont {Thibault}},\ }\href
  {https://doi.org/https://doi.org/10.1016/j.nuclphysa.2003.11.003} {\bibfield
  {journal} {\bibinfo  {journal} {Nuclear Physics A}\ }\textbf {\bibinfo
  {volume} {729}},\ \bibinfo {pages} {337} (\bibinfo {year} {2003})},\ \bibinfo
  {note} {the 2003 NUBASE and Atomic Mass Evaluations}\BibitemShut {NoStop}%
\bibitem [{\citenamefont {Lunney}\ \emph {et~al.}(2003)\citenamefont {Lunney},
  \citenamefont {Pearson},\ and\ \citenamefont {Thibault}}]{Lunney2003}%
  \BibitemOpen
  \bibfield  {author} {\bibinfo {author} {\bibfnamefont {D.}~\bibnamefont
  {Lunney}}, \bibinfo {author} {\bibfnamefont {J.~M.}\ \bibnamefont
  {Pearson}},\ and\ \bibinfo {author} {\bibfnamefont {C.}~\bibnamefont
  {Thibault}},\ }\href {https://doi.org/10.1103/RevModPhys.75.1021} {\bibfield
  {journal} {\bibinfo  {journal} {Rev. Mod. Phys.}\ }\textbf {\bibinfo {volume}
  {75}},\ \bibinfo {pages} {1021} (\bibinfo {year} {2003})}\BibitemShut
  {NoStop}%
\end{thebibliography}%

\end{document}